\documentclass{article}

\usepackage{arxiv}

\usepackage[utf8]{inputenc} 
\usepackage[T1]{fontenc}    
\usepackage{hyperref}       
\usepackage{url}            
\usepackage{booktabs}       
\usepackage{amsfonts}       
\usepackage{nicefrac}       
\usepackage{microtype}      
\usepackage{lipsum}

\usepackage{graphicx}
\usepackage{amssymb, amsmath, lineno, epsfig, bm, latexsym}
\usepackage{lscape} 
\usepackage{ulem}
\newtheorem{remark}{Remark}

\title{Semiparametric inference for the scale-mixture of normal partial linear regression model with censored data}

\author{
  Mehrdad Naderi \\
  Department of Statistics\\
  Faculty of Natural \& Agricultural Sciences\\
  University of Pretoria\\
  Pretoria, South Africa\\
  \texttt{m.naderi@up.ac.za} \\
   \And
  Elham Mirfarah\\
  Department of Statistics\\
  Faculty of Natural \& Agricultural Sciences\\
  University of Pretoria\\
  Pretoria, South Africa\\
  \texttt{mirfarah.elham@gmail.com} \\
   \AND
   Matthew Bernhardt \\
  Department of Statistics\\
  Faculty of Natural \& Agricultural Sciences\\
   University of Pretoria\\
   Pretoria, South Africa\\
   \And
   Ding-Geng Chen \\
   Department of Biostatistics\\
    University of North Carolina\\
     Chapel Hill, NC 27599, USA\\
   \texttt{dinchen@email.unc.edu} \\
}

\begin{document}
\maketitle

\begin{abstract}
In the framework of censored data modelling, the classical linear regression model that assumes normally 
distributed random errors has received increasing attention in recent years, mainly for mathematical 
and computational convenience. However, practical studies have often criticized this linear regression 
model due to its sensitivity to departure from the normality and from the partial nonlinearity. This paper proposes to solve 
these potential issues simultaneously in the context of the partial linear regression model by
assuming that the random errors follow a scale-mixture of normal (SMN) family of distributions. The proposed method allows us 
to model data with great flexibility, accommodating heavy tails and outliers. By implementing the B-spline function
and using the convenient hierarchical representation of the SMN distributions, a computationally analytical EM-type 
algorithm is developed to perform maximum likelihood inference of the model parameters.
Various simulation studies are conducted to investigate the finite sample properties as well as the robustness of 
the model in dealing with the heavy-tails distributed datasets. Real-word data examples are finally analyzed for illustrating
the usefulness of the proposed methodology.
\end{abstract}


\section{Introduction} \label{Intro}
Regression models form the basis for a large number of statistical inference procedures. The main purpose of 
regression analysis is to explore the relationship between a continuous response variable $Y$ and a $p$-dimensional 
covariate vector $\bm x\in\mathbb{R}^p$. More precisely, the regression models aim to find the expected value of 
$Y$ for a given level of covariate vector $\bm x$, say $E(Y|\bm x) = g(\bm x)$. These models can be found in a broad array of scientific fields, 
including econometrics, engineering and medical studies, and allow judgments to be made on data within these fields.
The parametric regression models in which the function $g(\cdot)$ can be specified in terms of a small number of parameters are widely used 
for data exploration since the parameters can be interpreted as the effects of covariates on the response variable. For instance, 
the classical linear regression model assumes that $Y=\bm{\beta}^\top\bm{x}+\epsilon,$ where $\epsilon$ represents the error term followed 
by a normal distribution and $\bm{\beta}^\top=(\beta_0,\beta_1,\ldots,\beta_{p-1})$. However, the nonlinearity (or partial nonlinearity), as well as invalid distribution assumptions, 
might increase the model misspecification and misleading inference. In this regard, the semiparametric techniques can provide 
an alternative platform for data analysis. One of the most acknowledged semiparametric models in the regression framework is the partially
linear regression (PLR) model. The PLR model assume that the response variable $Y$ not only has linear relationship with certain 
covariates but also is regressed by another covariate $z$ with unknown smooth function. More concretely, the PLR model allows both parametric 
and nonparametric specifications in the regression function and can be written as
\begin{equation}\label{PLR-Model}
Y=\bm{\beta}^\top \bm{x} +\psi(z)+ \epsilon.
\end{equation}
where $\psi(\cdot)$ is an unknown smooth function.
It can be observed from \eqref{PLR-Model} that the covariates are separated into parametric components $\bm x$ and nonparametric
component $z$. The parametric part of the model can be interpreted as a linear function, while the nonparametric part frees the 
rest of the model from any structural assumptions. As a result, the estimator of $\bm\beta$ is less affected by the model bias.
Since the introduction of PLR model by \cite{engle1986}, it has received attention in economics, social and biological 
sciences. See for instance \cite{Xue2006,Holland2017,Kim2018, He2019} and \cite{Taavoni2020} among others. Moreover, various alternative approaches to the penalized least-squares method, considered by \cite{engle1986}, were 
proposed for estimating the PLR models. For instance, \cite{robinson1988} exploited the profile least squares estimator for $\bm\beta$ 
and the Nadaraya-Watson kernel estimate \cite{Nadaraya1964,watson1964} for the unknown function. \cite{severini1994} also 
proposed a quasi-likelihood estimation method. However, to the best of our knowledge, none of the previous works consider specific 
distributions on the error term. 

In this article we propose three aspects. The first and main objective of this contribution is to develop a PLR model in which 
the error term is followed by an SMN distribution. The SMN family of distributions is an extension of the normal model with fat tails. 
It contains the student-$t$, slash, Laplace and contaminated normal as the special cases. Comprehensive surveys of the SMN family of 
distributions are available in \cite{meza2012,Garay2015} and \cite{garay2016}, among others. Our PLR model secondly implements the 
basis splines estimator as a powerful nonparametric approach of kernel estimation. As discussed in Section \ref{sec1}, the spline functions 
are piecewise polynomial functions where the weights in the sum are parameters that have to be estimated. The basis functions of the spline
will allow us to build a flexible model across the whole range of the data. The regression models with censored dependent variable 
have been considered in fields such as econometric engineering analysis, clinical essays, medical surveys, among others. Lastly our PLR model 
with B-spline consideration is enriched by assuming the interval-censoring scheme on the response variable, as an extension of 
\cite{Garay2015,garay2016}. 

The rest of the paper is therefore organized as follows. In Section \ref{sec1}, we provide a brief overview on the SMN family of 
distributions as well as B-spline functions. Section \ref{sec2} then presents the scale-mixture of normal partial linear regression model 
with interval-censored data, hereafter PLR-SMN-IC model. We also discuss on the implementation of a expectation-conditionally maximization either(ECME) algorithm, proposed by 
\cite{liu1994}, for obtaining the maximum-likelihood (ML) parameter estimates in Section \ref{sec2}. Simulation studies are conducted 
in Section \ref{sec4} to examine the properties of the proposed methodology. Finally the superiority of our model is illustrated in 
Section \ref{sec5} by analyzing two real-world datasets. Concluding remarks and possible directions for future research are 
discussed in Section \ref{sec6}.

\section{Background and notation}\label{sec1}
In this section we briefly review the SMN family of distributions and polynomial spline basis function.
For the sake of notation, let $\phi(\cdot;\mu,\sigma^2)$ and $\Phi(\cdot;\mu,\sigma^2)$ represent the 
probability density function (pdf) and cumulative distribution function (cdf) of a normal distribution 
with mean $\mu$ and variance $\sigma^2$, denoted by $\mathcal{N}(\mu,\sigma^2)$. 

\subsection{An overview on the scale-mixture of normal family of distributions}
Formally, the scale-mixture of normal (SMN) distribution is generated by scaling the variance of a normal
distribution with a positive weighting random variable $U$. More specifically, a random variable $V$ is 
said to have a SMN distribution, denoted by $\mathcal{SMN}(\mu, \sigma^2,\bm\nu)$, if it is generated by
the stochastic linear representation
\begin{equation}\label{SMN-lin-rep}
V=\mu+U^{-1/2}Z,\qquad Z\perp U,
\end{equation}
where $Z\sim \mathcal{N}(0,\sigma^2)$, $U$ is a scale-mixing factor with the cdf $H(\cdot;\bm\nu)$, indexed by the parameter
$\bm\nu$, and the symbol $\perp$ indicates independence. Referring to \eqref{SMN-lin-rep}, the hierarchical 
representation of the SMN distribution is
\begin{equation}\label{SMN-hir-rep}
V|U=u\sim N(\mu, u^{-1}\sigma^2), \qquad U\sim H(u; \bm\nu). \nonumber
\end{equation}
Accordingly, the pdf of the random variable $V$ can be expressed as
\begin{equation}\label{SMN-pdf}
f_{_\text{SMN}}(v;\mu, \sigma^2,\bm\nu) = \int_{0}^\infty \phi(v;\mu, u^{-1}\sigma^2)\; dH(u; \bm\nu),
\qquad v\in\mathbb{R}.
\end{equation}
In what follows, $f_{_\text{SMN}}(\cdot;\bm\nu)$ and $F_{_\text{SMN}}(\cdot;\bm\nu)$ will be used to denote the
pdf and cdf of the standard SMN distribution ($\mu=0,\sigma^2=1$). Depending on the choice of $H(\cdot;\bm\nu)$, 
a wide range of distributions can be generated using \eqref{SMN-pdf}. We focus on a few commonly used special
cases of the SMN family of distributions in this paper:

\begin{itemize}
	\item[$\bullet$] Normal (N) distribution: The SMN family of distributions contains the normal model as $U=1$ 
	with probability one.
	
	\item[$\bullet$] Student-$t$ (T) distribution: If $U\sim Gamma ({\nu}/{2},{\nu}/{2})$, where $Gamma(\alpha, 
	\beta)$ represents the  gamma distribution with shape and scale parameters $\alpha$ and $\beta$, respectively, 
	the random variable $V$  then follows the Student-$t$ distribution, $V\sim \mathcal{T}(\mu,\sigma^2,\nu)$. 
	For $\nu =1$ the Student-$t$ distribution turns into the Cauchy distribution which has no defined mean and 
	variance.
	
	\item[$\bullet$] Slash (SL) distribution: Let $U$ in \eqref{SMN-lin-rep} follows $Beta(\nu,1)$, where $Beta(a,b)$ 
	signifies the beta distribution with parameter $a$ and $b$. Then, $V$ distributed as a slash model, denoted by 
	$V\sim\mathcal{SL}(\mu,\sigma^2,\nu)$, with pdf
	\[
	f_{_\text{SL}}(v;\mu,\sigma^2,\nu) = \nu \int_{0}^1 u^{\nu-1}\phi(v;\mu, u^{-1}\sigma^2)\; du,
	\qquad v\in\mathbb{R}.
	\]
	
	\item[$\bullet$] Contaminated-normal (CN) distribution: Let $U$ be a discrete random variable with pdf
	\[
	h(u;\nu,\gamma)= \nu \mathbb{I}_\gamma(u)+(1-\nu)\mathbb{I}_1(u), \qquad \nu,\gamma\in (0,1),
	\]
	where $\mathbb{I}_A(\cdot)$ represents the indicator function of the set $A$. The random variable $V$ in
	\eqref{SMN-lin-rep} then follows the contaminated-normal distribution, $V\sim \mathcal{CN}(\mu,\sigma^2,\nu,\gamma)$,
	which has the pdf
	\[
	f_{_\text{CN}}(v;\mu,\sigma^2,\nu,\gamma)= \nu \phi(v; \mu,\gamma^{-1}\sigma^2)+(1-\nu)\phi(v; \mu,\sigma^2),
	\qquad v\in\mathbb{R}.
	\]
	Note that in the pdf of CN distribution,  the parameter $\nu$ denotes the proportion of outliers (bad points)
	and $\gamma$ is the contamination factor.
\end{itemize}
More technical details and information on the SMN family of distributions, used for the calculation of
some conditional expectations involved in the proposed EM-type algorithm, are provided in the Appendix A 
with proof in \cite{Garay2015}.

\subsection{B-spline function description}
The basis spline function, or B-spline in short, is a numerical tool that was originally introduced by \cite{curry1965} and recently 
received considerable attention in the statistical analysis of density estimation. For a given degree, smoothness and domain partition, 
the B-spline function provides a sophisticated approach to approximate the unknown function $\psi(\cdot)$. To approximate 
$\psi(x)$ on the interval $[a, b]$, any spline function of $d$ degree on a given set of $m$ interior knots, say 
$a=x_1<x_2<\cdots <x_{m+2}=b$, can formally be represented as a unique linear combination of B-splines:
\begin{equation}\label{BS-bais}
\psi(x) = \sum_{i=1}^{m+d} \alpha_i B_i^d(x),
\end{equation}
where the B-spline of $d$ degree is defined recursively by
\[
B_i^d(x)= \frac{x-x_i}{x_{i+d}-x_i}B_i^{d-1}(x) + \frac{x_{i+d+1} - x}{x_{i+d+1}-x_{i+1}}B_{i+1}^{d-1}(x), \qquad
B_i^0(x) = \mathbb{I}_{[x_i, x_{i+1})}(x),
\]
Although the piecewise linear approximation, case with $d =1$, is attractively simple, it produces a visible roughness, 
unless the knots $x_i$ are close to each other.

Theoretically, a B-spline function of degree $d$ with $m$ knots should be a $(d-1)$ continuously differentiable function, 
in each of the intervals $(a, x_2), (x_2, x_3),\ldots, (x_{m+1}, b)$. This condition may complicate the 
approximation issue. However, practical studies confirm that quadratic and cubic B-spline functions usually provide robust platforms
that have the minimal requirement, $\psi(x)$ should be twice continuously differentiable smooth function. Comprehensive review on the theory
of spline function can be found in \cite{schumaker2007}. 

In this paper, we consider the cubic B-spline function, i.e. $d=3$. The number of interior knots is also chosen by either $m_1=[n^{1/3}]+1$ 
or $m_2=[n^{1/5}]+1$, where $[a]$ denotes the largest integer smaller than $a$ and $n$ is the sample size. For the locations of knots, two scenarios
are considered: the equally-spaced (ES), and the equally-spaced quantile (ESQ). As investigated in our simulation studies, 
our strategy works well under these assumptions. However, for the practical studies if a large number of knots is required 
and there is not enough data on the boundary, we may need to obtain the knots through an unequally spaced technique to avoid 
singularity problems \cite{Huang2006}.

\section{Proposed methodology}\label{sec2}
\subsection{Model formulation}
In this section, we consider the partial linear regression model where the random error follows an SMN family of 
distributions. Let $\bm Y=\{Y_1,\ldots,Y_n\}$ be a set of response variables and $\bm x_i$ a vector of explanatory variable
values corresponding to $Y_i$. The PLR model based on the SMN distribution is defined as
\begin{equation}\label{PLR}
Y_i = \bm{\beta}^\top \bm{x}_i + \psi(z_i) + \epsilon_{i}, \qquad \epsilon_{i}\stackrel{iid}{\sim}\mathcal{SMN}(0,\sigma^2,\bm\nu), 
\qquad i=1,2,\ldots,n,
\end{equation}
where $iid$ represents the independent and identically distributed, $z_i$ is a univariate covariate such as the confounding factor,
and $\psi(\cdot)$ is an unknown smooth function playing the role of the nonparametric component. Clearly we can conclude that
$Y_i\stackrel{iid}{\sim}\mathcal{SMN}(\bm{\beta}^\top \bm{x}_i + \psi(z_i),\sigma^2,\bm\nu)$. 

Censored time-to-event data is widely seen in health studies, such as, time-to-death in cancer research, time-to-infection in HIV, TB and COVID-19 studies. Among all the censored data, the interval-censored data is the most general type of data, which covers the typical right-censored and left-censored data as special cases.  Interval-censored data can also be generated from other science fields, such as detection limits of quantification in environmental, toxicological and pharmacological studies. Therefore, we focus on interval-censored data in this paper. We assume that the set of joint variables $\{W_i,\rho_i\}$ are observed where $W_i$ represents
the uncensored reading ($W_i = Y_{i}^o$) or interval-censoring ($W_i =(C_{i1},C_{i2})$) and $\rho_i$ is the censoring indicator:
$\rho_i=1$ if $C_{i1} \leq Y_i \leq C_{i2}$ and $\rho_i=0$ if $Y_i = Y_{i}^o$. Note that in this setting if $C_{i1} =-\infty$ 
(or $C_{i2} = +\infty$) the left-censoring (or right-censoring) is occurred and in the case $-\infty\neq C_{i1}<C_{i2} \neq +\infty$ 
the interval-censored realization is observed. We establish our methodology based on the interval-censoring scheme, however, 
the left/right-censoring schemes are also investigated in the simulation and real-data analyses. 
We will refer to the PLR model of censored data based on the special cases of the SMN family of distributions as PLR-N-IC, PLR-T-IC, 
PLR-SL-IC and PLR-CN-IC for the N, T, SL and CN cases, respectively. 

For ease of exposition and based on the aforementioned setting, one can divide $\bm Y$ to the sets of observed responses, $\bm Y^o$, 
and censored cases. We can therefore view $\bm Y$ as the latent variable since it is partially unobserved. 
Under these assumptions, the log-likelihood function for $\bm\Theta=(\bm\beta^\top,\sigma^2,\bm\nu)$ of the PLR-SMN-IC model 
can be written as
\begin{eqnarray} \label{loglike}
\ell(\bm\Theta|\bm w, \bm\rho) = \sum_{i=1}^n\left[{\sigma}^{-1}
f_{_\text{SMN}}\left(\frac{w_i-\mu_i}{\sigma};\bm\nu\right)\right]^{1-\rho_i} 
\left[ F_{_\text{SMN}}\left(\frac{c_{i2}-\mu_i}{\sigma}; \bm\nu\right) - 
F_{_\text{SMN}}\left(\frac{c_{i1}-\mu_i}{\sigma}; \bm\nu\right)\right]^{\rho_i} ,
\end{eqnarray}
where $\mu_i=\bm{\beta}^\top \bm{x}_i+\psi(z_i)$, $\bm\rho = (\rho_1,\ldots,\rho_n)^\top$ are the censoring indicators, and $\bm w=(w_1,\ldots,w_n)^\top$ denote the 
realizations of $\bm W=(W_1,\ldots,W_n)^\top$.

Due to complexity of the log-likelihood function \eqref{loglike}, there is no analytical solution to obtain the ML estimates of parameters
and the smooth function. A numerical search algorithm should therefore be developed. With the embedded hierarchical representation 
\eqref{SMN-hir-rep} and B-spline function, an innovative EM-type algorithm is developed to calibrate the PLR-SMN-IC model to the
data.

\subsection{Estimation via an EM-type algorithm}
In this section, an EM-type algorithm is developed for calibrating the PLR-SMN-IC model to the data. In order to do this, we first
replace the basis expansion of $\psi(z_i)$ defined in \eqref{BS-bais} into the model \eqref{PLR}. Immediately, we can obtain 
\begin{equation}\label{PLR2}
Y_i = \bm{\beta}^\top \bm{x}_i + \sum_{j=1}^{m+d} \alpha_j B_j^d(z_i) + \epsilon_{i}= \bm{\beta}^\top \bm{x}_i + 
\bm\alpha^\top \bm B^d(z_i) + \epsilon_{i},\qquad \epsilon_{i}\stackrel{iid}{\sim}SMN(0,\sigma^2,\bm\nu),
\end{equation}
where $\bm\alpha=(\alpha_1, \ldots,\alpha_{m+d})^\top$, $\bm B^d(z)=(B_1^d(z),\ldots, B_{m+d}^d(z))^\top$. For convenience, 
the obtained model in \eqref{PLR2} can be rewritten as
\begin{equation}\label{PLR3}
Y_i = \tilde{\bm\beta}^\top \tilde{\bm x}_i + \epsilon_{i}, \qquad \epsilon_{i}\stackrel{iid}{\sim}SMN(0,\sigma^2,\bm\nu), \nonumber
\end{equation}
where $\tilde{\bm x}_i=(\bm x_i^\top,\bm B^{d^\top}(z_i))^\top$ are the pseudo covariates and $\tilde{\bm\beta}=
(\bm\beta^\top, \bm\alpha^\top)^\top$ is the pseudo parameter. We therefore have, 
$Y_i\sim \mathcal{SMN}(\tilde{\bm\beta}^\top \tilde{\bm x}_i, \sigma^2,\bm\nu), \text{ for } i=1,\ldots,n.$
Now using \eqref{SMN-hir-rep}, the hierarchical representation of the PLR-SMN-IC model is
\begin{align*}
Y_i |(\tilde{\bm x}_i, U=u_i) &\sim \mathcal{N}(\tilde{\bm\beta}^\top \tilde{\bm x}_i, u_i^{-1}\sigma^2), \\
U_i &\sim H(\cdot;\bm\nu).
\end{align*}
For the realizations $\bm y=(y_1,\ldots,y_n)^\top$ and the latent values $\bm u=(u_1,\ldots,u_n)^\top$, the log-likelihood 
function for $\bm \Theta=(\tilde{\bm\beta}^\top,\sigma^2,\bm\nu)$ associated with the complete data $\bm y_c=(\bm w^\top,\bm\rho^\top,\bm y^\top, \bm u^\top)^\top$, 
is therefore given by
\begin{equation}\label{log-com}
\ell_c(\bm\Theta|\bm y_c)= c-\frac{n}{2}\log \sigma^2 + \sum_{i=1}^n \log h(u_i;\bm \nu) - \frac{1}{2\sigma^2} \sum_{i=1}^n u_{i} (y_i-\tilde{\bm\beta}^\top \tilde{\bm x}_i)^2 ,
\end{equation}
where $h(\cdot;\bm \nu)$ is the pdf of $U_i$ and $c$ is an additive constant.

We then develop an expectation conditional maximization either (ECME; \cite{liu1994}) algorithm to estimate parameters 
from the PLR-SMN-IC model. As an extension of expectation conditional maximization (ECM; \cite{meng1993})
the ECME algorithm has stable properties (e.g. monotone convergence and implementation simplicity) and can be implemented 
faster than ECM. The ML parameter estimates via ECME algorithm are obtained by maximizing the constrained $Q$-function with 
some CM-steps that maximize the corresponding constrained actual marginal likelihood function, called CML-steps.
The ECME algorithm for ML estimation of the PLR-SMN-IC model proceeds as follows:
\begin{itemize}
	\item[$\bullet$] {\bf Initialization:}
	Set the number of iteration to $k = 0$ and choose a relative starting point. 
	
	\item[$\bullet$] {\bf E-Step:} At iteration $k$, the expected value of the complete-data log-likelihood function
	\eqref{log-com}, known as the $Q$-function, is computed as
	\begin{align}\label{qfunc}
	Q(\bm\Theta| \hat{\bm\Theta}^{(k)})
	=& c-\frac{n}{2}\log \sigma^2+ \sum_{i=1}^n\hat{\Upsilon}_{i}^{(k)} -\frac{1}{2\sigma^2}\sum_{i=1}^n \left( \widehat{uy^2}_{i}^{(k)}+
	(\tilde{\bm\beta}^\top \tilde{\bm x}_i)^2 \hat{u}_{i}^{(k)} -2  \widehat{uy}_{i}^{(k)} \tilde{\bm\beta}^\top \tilde{\bm x}_i\right),
	\end{align}
	where $\widehat{uy^2}_{i}^{(k)} = E(U_{i}Y_i^2| w_i, \rho_i, \hat{\bm\Theta}^{(k)} ),$
	$\hat{u}_{i}^{(k)}  = E(U_{i} |  w_i, \rho_i, \hat{\bm\Theta}^{(k)} ),$ 
	$\widehat{uy}_{i}^{(k)} = E(U_{i}Y_i |  w_i, \rho_i, \hat{\bm\Theta}^{(k)}),$ and
	$\hat{\Upsilon}_{i}^{(k)} = E\big(\log h(U_i;\bm\nu)|w_i,\rho_i, \hat{\bm\Theta}^{(k)}\big)$.
	In what follows, we discuss the computation of conditional expectations for both uncensored and censored
	cases. 		
	\begin{itemize}
		\item[(i)] For the uncensored observations, we have $\rho_i =0$ and so,
		$\hat{u}_{i}^{(k)}  = E(U_{i} |  Y=y_i, \hat{\bm\Theta}^{(k)})$,
		$\widehat{uy}_{i}^{(k)} = y_i\hat{u}_{i}^{(k)},$
		$\widehat{uy^2}_{i}^{(k)} = y_i^2 \hat{u}_{i}^{(k)},$ and	$
		\hat{\Upsilon}_{i}^{(k)} = E\big(\log h(U_i;\bm\nu_j)|Y=y_i, \hat{\bm\Theta}^{(k)} \big).$
		
		\item[(ii)] For the censored case in which $\rho_i =1$, we have
		\begin{align*}
		\hat{u}_{i}^{(k)}  & = E(U_{i} |  c_{i1} \leq Y_i \leq c_{i2}, \hat{\bm\Theta}^{(k)} ), \qquad
		\widehat{uy^2}_{i}^{(k)} = E(U_{i}Y_i^2|  c_{i1} \leq Y_i \leq c_{i2}, \hat{\bm\Theta}^{(k)} ), \\
		\widehat{uy}_{i}^{(k)} & =
		E( U_{i}Y_i |  c_{i1} \leq Y_i \leq c_{i2}, \hat{\bm\Theta}^{(k)} ),\qquad
		\hat{\Upsilon}_{i}^{(k)} = E\big(\log h(U_i;\bm\nu_j)|c_{i1} \leq Y_i \leq c_{i2}, \hat{\bm\Theta}^{(k)} \big).
		\end{align*}
	\end{itemize}
	The closed form of the conditional expectations for the particular cases of the SMN family of distributions 
	are provided in Appendix A. 
	
	\item[$\bullet$] {\bf CM-step:} The $M$-step consists of maximizing the $Q$-function with respect to $\bm\Theta^{(k)}$.
	The maximization of \eqref{qfunc} over $\tilde{\bm\beta}$ and  $\sigma^2$ lead to the following CM estimators:
	\begin{align*}
	\hat{\tilde{\bm\beta}}^{(k+1)}&= \left( \sum_{i=1}^n \hat{u}_{i}^{(k)} \tilde{\bm x}_i \tilde{\bm x}_i^\top \right)^{-1}
	\sum_{i=1}^n \widehat{uy}_{i}^{(k)} \tilde{\bm x}_i,\\
	\hat{\sigma}^{2(k+1)} &=\frac{1}{n} \sum_{i=1}^n
	\left(\widehat{uy^2}_{i}^{(k)}-2 \widehat{uy}_{i}^{(k)} \hat{\tilde{\bm\beta}}^{(k+1)^\top }\tilde{\bm x}_i+\hat{u}_{i}^{(k)}
	\left( \hat{\tilde{\bm\beta}}^{(k+1)^\top}\tilde{\bm x}_i \right)^2 \right).
	\end{align*}
	\item[$\bullet$] {\bf CML-step:} The update of $\bm\nu$ crucially depends on the conditional expectation $\hat{\Upsilon}_{i}^{(k)}$
	which is quite complicated. However, we can update $\bm\nu$ through maximizing the actual log-likelihood function as
	\begin{eqnarray}\label{nuupdate}
	\hat{\bm\nu}^{(k+1)} &=& \arg\max_{\bm\nu} \sum_{i=1}^n (1-\rho_i) \log\left[
	f_{_\text{SMN}}\left(\frac{w_i-\hat{\tilde{\bm\beta}}^{(k+1)^\top }\tilde{\bm x}_i}{\hat{\sigma}^{(k+1)}}; 
	\bm\nu\right)\big/\hat{\sigma}^{(k+1)}\right] \nonumber \\
	&& \qquad \qquad + \rho_i\log\left[ F_{_\text{SMN}}\left(\frac{c_{i2}-\hat{\tilde{\bm\beta}}^{(k+1)^\top }\tilde{\bm x}_i}{\hat{\sigma}^{(k+1)}}; \bm\nu\right) - 
	F_{_\text{SMN}}\left(\frac{c_{i1}-\hat{\tilde{\bm\beta}}^{(k+1)^\top }\tilde{\bm x}_i}{\hat{\sigma}^{(k+1)}};\bm\nu\right)\right].
	\end{eqnarray}
	The $\texttt{R}$ function $\mathbf{nlminb}(\cdot )$	is used to update $\bm\nu$ in the numerical parts of the current paper.
\end{itemize}
\begin{remark}\label{rem1}
	To facilitate the update of $\bm\nu=(\nu,\gamma)$ for the PLR-CN-IC model in the above ECME algorithm, 
	one can introduce an extra latent binary variable $B_{i}$ such that $B_{i}=1$ if an observation $y_i$ is a bad point (outlier) and 
	$B_{i}=0$ if $y_i$ is a good point. The hierarchical representation of the PLR-CN-IC model can therefore be written as
	\begin{align}\label{CNup}
	Y_i |(\tilde{\bm x}_i, U=u_i,B_{i}=1)\sim \mathcal{N}(\tilde{\bm\beta}^\top \tilde{\bm x}_i, u_i^{-1}\sigma^2),\quad
	U_i| (B_{i}=1)\sim h(\cdot;\nu,\gamma),\quad
	B_{i}\sim \mathcal{B}(1,\nu),
	\end{align}
	where $\mathcal{B}(1,\nu)$ denotes the Bernoulli distribution with success probability $\nu$. 
	Consequently, by computing the $Q$-function
	based on \eqref{CNup}, the update of $\nu$ is $\hat{\nu}^{(k+1)} ={n}^{-1} {\sum_{i=1}^n\hat{b}_{i}^{(k)}},$
	where
	\begin{equation*}
	\hat{b}_{i}^{(k)} = \left\lbrace \begin{array}{lll}
	\dfrac{\hat{\nu}^{(k)} \phi\big(y_i;\hat{\tilde{\bm\beta}}^{(k)^\top }\tilde{\bm x}_i, \hat{\gamma}^{-1(k)}\hat{\sigma}^{2(k)}\big)}{\hat{\nu}^{(k)} \phi\big(y_i;\hat{\tilde{\bm\beta}}^{(k)^\top }\tilde{\bm x}_i, \hat{\gamma}^{-1(k)}\hat{\sigma}^{2(k)}\big)+(1-\hat{\nu}^{(k)}) \phi\big(y_i; \hat{\tilde{\bm\beta}}^{(k)^\top }\tilde{\bm x}_i, \hat{\sigma}^{2(k)}\big)},\qquad
	\text{for the uncensoed cases,}\\
	\\
	\dfrac{\hat{\nu}^{(k)} \Big(\Phi\big(c_{i2};\hat{\tilde{\bm\beta}}^{(k)^\top }\tilde{\bm x}_i,\hat{\gamma}^{-1(k)} \hat{\sigma}^{2(k)}\big)-\Phi\big(c_{i1};\hat{\tilde{\bm\beta}}^{(k)^\top}\tilde{\bm x}_i,\hat{\gamma}^{-1(k)} \hat{\sigma}^{2(k)}\big)\Big)}{F_{CN}\big(c_{i2};\hat{\tilde{\bm\beta}}^{(k)^\top }\tilde{\bm x}_i, \hat{\sigma}^{2(k)},\hat{\nu}^{(k)} ,\hat{\gamma}^{(k)}\big)-F_{CN}\big(c_{i1};\hat{\tilde{\bm\beta}}^{(k)^\top }\tilde{\bm x}_i, \hat{\sigma}^{2(k)},\hat{\nu}^{(k)} ,\hat{\gamma}^{(k)}\big)},
	\quad \text{for the censoed cases.}
	\end{array}
	\right. 
	\end{equation*}
	Since there is no closed-form solution for $\hat{\gamma}^{(k+1)}$, we update $\gamma$ by maximizing the constrained actual 
	observed log-likelihood function \eqref{nuupdate} as a function of $\gamma$.
\end{remark}

\subsection{Computational aspects}
\subsubsection{Initial values}
The choice of starting points plays a critical role speeding up parameter estimation via the EM-type algorithm and to guarantee
reaching stationarity in the ML solutions. As a convenient approach to generate sensible initial values, we fit a classical linear model to 
data and find the estimate of $\bm\beta$. To do this, the $\texttt{R}$ function ``$\mathbf{lm}(\cdot)$" is used. We also set $\sigma^2$ as the
average of squared residuals of the classical linear model. The obtained parameter estimates of $\bm\beta$ and $\sigma^2$ are used as the initial points for 
implementing PLR-N-IC model. By calibrating the PLR-N-IC model to the data, the parameter estimates are exploited as the starting 
values for the PLR-T-IC, PLR-SL-IC, PLR-CN-IC models. We adapt the scale mixture factor parameter $\hat{\bm\nu}^{(0)}$ so that it corresponds 
to an initial assumption near normality. For instance, we set $\hat{\nu}^{(0)}=20$ in the PLR-T-IC, PLR-SL-IC models.

\subsubsection{Convergence}
The process of the EM algorithm can be iterated until a suitable convergence rule is satisfied. Herein, 
the incremental likelihood property of EM-type algorithm is used to detect the convergence. We terminate the algorithm 
either when the maximum number of iterations $K_{\max}=2000$ has been reached or 
\[
\frac{\ell(\bm{\hat{\Theta}}^{(k+1)})-\ell(\bm{\hat{\Theta}}^{(k)})}{|\ell(\bm{\hat{\Theta}}^{(k)})|}\leq\varepsilon,
\]
where $\ell(\cdot)$ is the log-likelihood function defined in \eqref{loglike} and $\varepsilon$ is a user specified tolerance. 
In our study, the tolerance $\varepsilon$ is considered as $10^{-5}$.

\subsubsection{Model Selection}
The models in competition in our data analysis are compared using the most commonly 
used information-based measures. Following \cite{Huang2006}, we vary the number of knots in a
relatively large range and choose the one which minimizes the Akaike information criterion (AIC) or 
Bayesian information criteria (BIC) defined as
\[
\text{AIC} = 2(m+d+p+s+1)-2\ell_{\max}\quad \text{and}\quad \text{BIC} = (m+d+p+s+1)\log n - 2\ell_{\max},
\]
where $m, d, p$ and $s$ denote, respectively, the number of knots, degree of the spline, number of covariates and number 
of parameters of the scale mixing factor $U$, and $\ell_{\max}$ is the maximized log-likelihood value. Models with lower 
values of AIC or BIC are considered more preferable. It should be noted that $m$ ranging from 3 to 10 are usually adequate and 
the results are quite stable when we vary the number of knots.

\section{Simulation studies}\label{sec4}
In this section, four Monte-Carlo simulation studies are conducted in order to verify the asymptotic properties of the ML estimates, 
to assess the fitting performance of the model, and to check the robustness of the proposed model in dealing with highly peaked, 
heavily tailed data as well as its robustness in presence of outliers. For the sake of data generation, it should be noted that
one of the simplest ways of interval-censored data generation is to consider $C_{i1} = Y_i - U_i^{(1)}$ and 
$C_{i2} = Y_i + U_i^{(2)}$ where $U_i^{(1)}$ and $U_i^{(2)}$ are two independent continuous variables followed by $\mathcal{U}(0, c)$ 
such that the non-informative condition (1.2) of \cite{gomez2009} is fulfilled. Here $\mathcal{U}(a,b)$ represents the 
uniform distribution on interval $(a,b)$. Recommended by \cite{gomez2009}, a way to go around non-informative condition is to construct 
$C_{i1} = \max( Y_i - U_i^{(1)}, Y_i + U_i^{(2)} -c)$ and $C_{i2} = \min(Y_i + U_i^{(2)}, Y_i - U_i^{(1)}+c)$ with $c = 1$,
which can be shown that fulfills the non-informative condition. Let $\bm y=(y_1,\ldots,y_n)^\top$ be the $n$ realizations from 
model \eqref{PLR}. To have a $p\%$ interval-censored dataset, the following steps are used in our simulation studies. 
\begin{itemize}
	\item[] 1) Compute the number of censored samples $\mathcal{NC}= [n \times p]+1$ and then, generate an index set, $\mathcal{IND}$,
	as a sample of size $\mathcal{NC}$ from $\{1, 2, \cdots, n\}$ without replacement by using the $\mathtt{R}$ function $sample(\ )$.
	\item[] 2) For $i= 1, \ldots,n$, if $i \in \mathcal{IND}$, we then generate two independent random variables, $U_i^1$ and $U_i^2$, 
	independently from $\mathcal{U}(0,c)$. Finally, we have  
	$C_{i1} = \max( Y_i - U_i^{(1)}, Y_i + U_i^{(2)} -c), C_{i2} = \min(Y_i + U_i^{(2)}, Y_i - U_i^{(1)}+c) $.
\end{itemize}

\subsection{Finite sample properties of the ML estimates}
To assess the performance of the ECME algorithm for obtaining ML estimations, we conduct an extensive simulation study 
under various scenarios. We generate the response variable from the following model
\[
Y_i = \bm{\beta}^\top \bm{x}_i + \psi(z_i) + \epsilon_{i}, \qquad i=1,2,\ldots,n,
\]
where $\epsilon_{i}$ is generated from either the N, T, SL or the CN distributions for various sample sizes ranging from 50
to 800, $\psi(z)=\exp(z/3) - 1$, and $\bm\beta=(1,2,-2)$. We also consider $\bm x_i^\top=(x_{i1},x_{i2},x_{i3})$, where $x_{i1}\sim\mathcal{N}(0,1)$,
$x_{i2}\sim\mathcal{B}(1,0.5)$, $x_{i3}\sim\mathcal{U}(-4,1)$ and $z_i\sim \mathcal{U}(-1,2)$.
Furthermore, we set $\sigma^2 = 2$, $\nu = 3$ for the T and SL distributions and $\nu=0.4$ and $\gamma=0.3$ for the CN model.
The considered levels of interval-censoring are 7.5\% and 30\%.

\begin{table}
	\centering
	\caption{The mean of model selection criteria (AIC and BIC) for various sample size and two approaches of the number of knots. \label{Tab1}}
	\hspace{-0.3cm}\scalebox{0.89}
	{\begin{tabular}{lcccccccccccccccccccc}
			\hline
			&&&& \multicolumn{2}{c}{PLR-N-IC} && \multicolumn{2}{c}{PLR-T-IC} 
			&& \multicolumn{2}{c}{PLR-CN-IC} && \multicolumn{2}{c}{PLR-SL-IC}\\
			\cline{5-6}\cline{8-9}\cline{11-12}\cline{14-15}
			Cens.       &  &  $n$ && $m_1$ & $m_2$ && $m_1$ & $m_2$ && $m_1$ & $m_2$ && $m_1$ & $m_2$\\
			\hline
			&   		&50	    && 189.252 & 188.859 && 221.472 & 221.504 && 251.299 & 251.435 && 209.108 & 209.038\\
			&&100	&& 372.238 & 370.656 && 438.647 & 437.863 && 497.653 & 497.330 && 412.790 & 410.142\\
			&AIC    	&200	&& 736.095 & 731.455 && 865.569 & 859.853 && 982.907 & 985.717 && 817.52 & 808.516\\
			&&400	&& 1453.720& 1452.680&& 1719.688& 1708.525&& 1947.138& 1956.133&& 1618.665 & 1605.993\\
			7.5\%		    &&800	&& 2862.746& 2890.951&& 3430.273& 3406.174&& 3864.258& 3896.971&& 3240.292 & 3199.743\\ \\
			&   		&50	    && 210.284 & 207.978 && 244.416 & 242.536 && 276.155 & 274.379 && 232.052 & 230.070\\
			&&100	&& 403.500 & 396.708 && 472.514 & 466.520 && 534.128 & 528.592 && 446.657 & 438.799\\
			&BIC        &200	&& 778.943 & 764.437 && 911.746 & 896.134 && 1032.381& 1025.297&& 863.696 & 844.797\\
			&&400   && 1513.671& 1496.586&& 1783.551& 1756.422&& 2014.992& 2008.021&& 1682.528& 1653.890\\
			&&800   && 2942.385& 2942.482&& 3514.596& 3462.390&& 3953.266& 3957.871&& 3324.615& 3255.959\\
			\hline
			&   		&50	    && 202.223 & 198.278 && 222.721 & 223.604 && 270.302 & 269.439 && 222.132 & 220.470\\
			&&100	&& 392.823 & 386.803 && 446.061 & 440.721 && 534.041 & 536.024 && 434.102 & 432.319\\
			&AIC    	&200	&& 768.888 & 777.524 && 873.019 & 862.663 && 1052.913& 1070.035&& 868.217 & 857.214\\
			&&400	&& 1510.605& 1532.380&& 1733.630& 1716.690&& 2053.333& 2120.194&& 1743.172& 1710.321\\
			30\%		    &&800	&& 2935.814& 3054.148&& 3472.984& 3407.593&& 4090.309& 4268.359&& 3474.358& 3418.554\\ \\
			&   		&50	    && 223.255 & 217.399 && 245.665 & 244.636 && 295.158 & 292.383 && 245.076 & 241.502\\
			&&100	&& 424.114 & 412.854 && 479.928 & 470.377 && 570.513 & 567.286 && 467.969 & 460.976\\
			&BIC        &200	&& 811.766 & 810.807 && 919.195 & 898.944 && 1102.387& 1109.615&&  914.393& 893.495\\
			&&400   && 1570.476& 1576.286&& 1797.493& 1764.587&& 2121.188& 2172.082&& 1807.035& 1758.219\\
			&&800   && 3015.453& 3105.677&& 3557.307& 3463.809&& 4179.317& 4329.259&& 3558.681& 3477.770\\
			\hline
	\end{tabular}}
\end{table}

In each replication of 500 trials, we fit the special cases of the PLR-SMN-IC model to the data under both assumptions of the 
number of interior knots $m_1$ and $m_2$ explained in Section \ref{sec1}. The locations of knots are also chosen based on the ESQ
scenario, even though some results are observed based on the ES method. Table \ref{Tab1} depicts the average values of the
AIC and BIC across all generated samples. As can be expected, the values of AIC and BIC are increased by exceeding the percentage 
of censoring. Results in Table \ref{Tab1} suggest that the chosen optimal number of interior knots depends on the level of censoring 
and the considered model. For instance, the BIC values of the PLR-T-IC model highlight the outperformance of the $m_2$ approach.
On the other hand, for the PLR-CN-IC model, one can conclude that the optimal number of interior knots for a 7.5\% censoring level is obtained via $m_1$
and for 30\% via $m_2$. The number of outperformance of the $m_2$ approach is however significantly higher than $m_1$ in this simulation
study. We therefore only focused on the $m_2$ approach in what follows of this simulation to shorten the length of the paper.

To investigate the estimation accuracies, we compute the bias and the mean squared error (MSE): 
\[
\text{BIAS}=\frac{1}{500}\sum_{j=1}^{500}(\hat{\theta}_j - \theta_{true})\qquad
\text{and} \qquad \text{MSE}=\frac{1}{500}\sum_{j=1}^{500}(\hat{\theta}_j - \theta_{true})^2,
\]
where $\hat{\theta}_j $ denotes the estimate of a specific parameter at the $j$th replication. In addition, we are interested in 
examining the accuracy of the $\psi(z)$ estimation in terms of the averaged integrated absolute bias (IABIAS) and mean 
integrated square error (MISE) defined as
\[
\text{IABIAS}= \frac{1}{500}\sum_{j=1}^{500}\left(\frac{1}{n}\sum_{i=1}^{n}\big|\widehat{\psi}_j(z_i) - {\psi}(z_i)\big| \right) \qquad
\text{and} \qquad \text{MISE}=\frac{1}{500}\sum_{j=1}^{500}\left(\frac{1}{n}\sum_{i=1}^{n}(\widehat{\psi}_j(z_i) - {\psi}(z_i))^2 \right),
\]
where $\widehat{\psi}_j(z_i)=\hat{\bm\alpha}_{j}^\top \bm B^3(z_i),$ is the estimate of $\psi(\cdot)$ in the $j$th replication.

\begin{figure}[!t]
	\centerline{\includegraphics[height=15.5cm, width=11cm, scale=0.5, angle=-90]{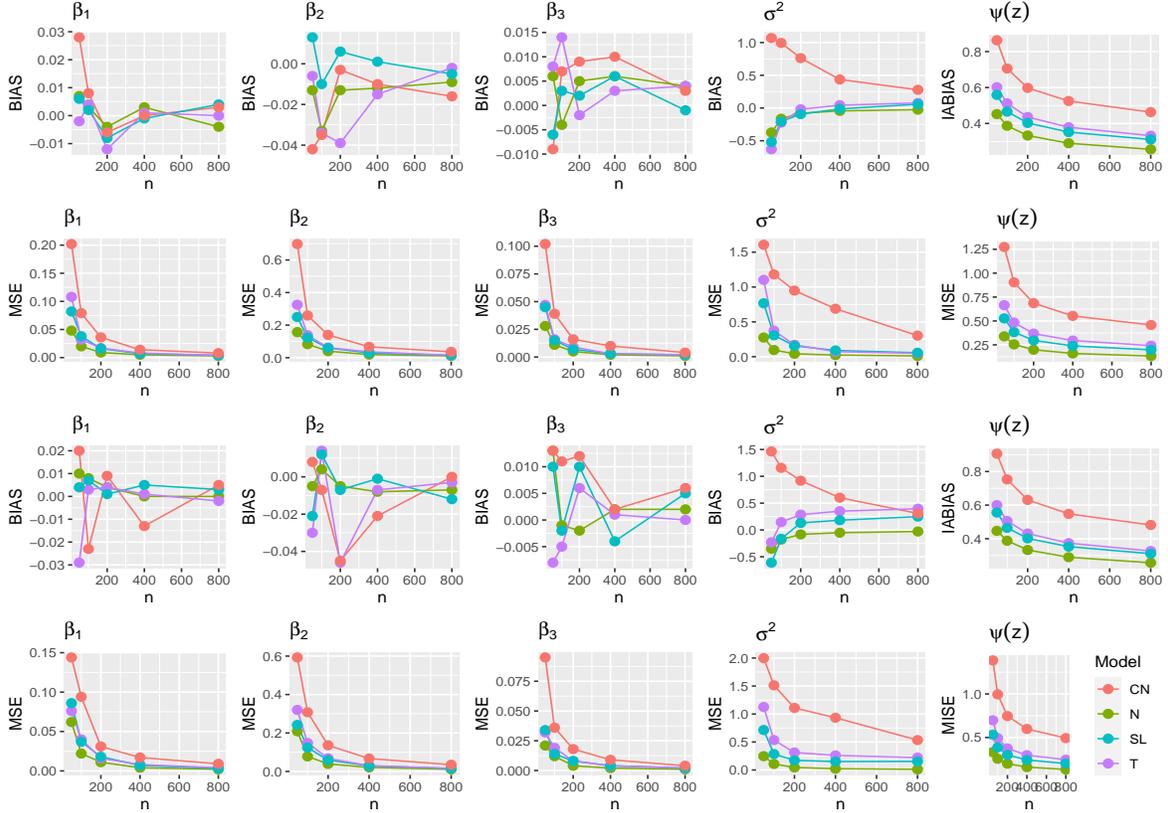}}\vspace{-0.3cm}
	\caption{Average BIAS and MSE of parameter estimates as well as average IABIAS and MISE of the estimated $\psi(z)$ in the
		PLR-SMN-IC models for two levels of censoring: 7.5\% (first and second rows from top) and 
		30\% (first and second rows from bottom).\label{sim1fig}}
\end{figure}

Figure \ref{sim1fig} plots the BIAS and MSE of the regression parameters ($\bm\beta$), $\sigma^2$ and the IABIAS and MISE
of the estimated $\psi(z)$ as a function of sample sizes for two levels of censoring. It can be observed that the 
estimates of the $\beta_i$s have very small (around zero) BIAS for all sample sizes. 
Moreover, as $n$ increases the BIAS of $\sigma^2$ and the IABIAS of the estimated $\psi(z)$ tend 
to zero. The plots in Figure \ref{sim1fig} also reveal that the MSE of the parameters and MISE of the estimated $\psi(z)$
tend to zero when the sample size is increased. These results indicate that our estimator of the regression parameters and
the estimation of $\psi(z)$ is rather accurate.

\subsection{Model comparison}

In this Monte-Carlo simulation experiment, the flexibility of the proposed PLR-SMN-IC model to modeling data generated 
form the some other distributions, is investigated. We generate $200$ samples of size 400 from the right-censored
linear regression model, 
\[
Y = \beta_0+\beta_1 x_1 +\beta_2 x_2 +\beta_3 x_3 + \sin(z\pi)+\epsilon
\]
where $\bm\beta^{\top}=(\beta_0,\beta_1,\beta_2,\beta_3) = c(1, 2, -2, 1)$, $\sigma^2 = 2$, $x_1\sim\mathcal{N}(0, 1)$, 
$x_2\sim\mathcal{B}(1,0.5)$, $x_3\sim\mathcal{U}(-2, 2)$ and $z\sim\mathcal{U}(-1, 7)$. The considered levels of censoring
are 7.5\%, 15\%, and 30\%. We generate the error terms under 
the three different scenarios from the SMN representation \eqref{SMN-lin-rep}. In the first scenario the random variable
$U$ in representation \eqref{SMN-lin-rep} is generated from the exponential distribution with mean 2, where as the 
Birnbaum-Saunders (BS) and generalized inverse Gaussian (GIG) models are used for the second and third scenarios, respectively.
The parameters are set to $\beta=1$ (scale parameter) and $\nu=2$ (shape parameter) for the BS distribution, and
to $(\kappa,\chi,\psi)=(-0.5,1, 3)$ for the GIG model.
Since the error term under these scenarios follows the Laplace, BS and GIG scale-mixture distributions, the 
notations PLR-L-IC, PLR-SBS-IC and PLR-SGIG-IC are used to denote the PLR model under these models.    	
Bare in mind that the considered non-normal data generation offers the desired level of leptokurtosis.
We also note that the PLR-L-IC, PLR-SBS-IC and PLR-SGIG-IC models are not considered in this paper since their conditional expectations 
involved in the ECME algorithm are not exist.

For each sample, we fit the PLR-N-IC, PLR-T-IC, PLR-SL-IC and PLR-CN-IC models to the data by assuming both $m_1$ and $m_2$
methods for choosing the number of knots, as well as ES and ESQ methods for the knots position. Table \ref{Tab2} depicts the 
average values of AIC and BIC measures over the 200 trials. Not surprisingly, under different levels of censoring, number of 
knots and position of the knots, all criteria favour PLR censored models based on heavy tail distributions. As highlighted in 
Table \ref{Tab2}, the PLR-T-IC model is the best in almost all cases. Furthermore, the outputs in this table suggest that $m_2$ preforms significantly better than $m_1$. However, it can be seen that there is no large difference in the AIC 
and BIC values between the two methods of knot position for all models, indicating their robustness against the position of knots.

\subsection{Imputation of censored observations in presence of noisy points}

\begin{figure}[!t]
	\centerline{\includegraphics[height=6cm, width=8cm]{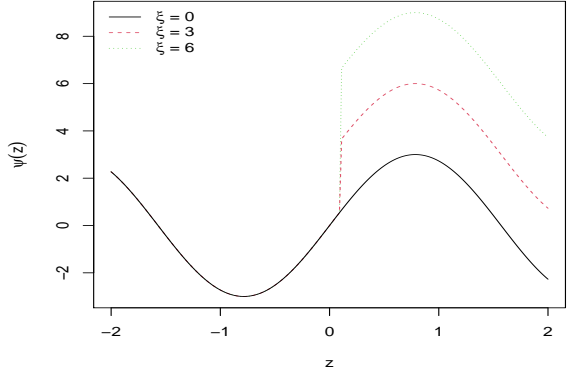}}\vspace{-0.3cm}
	\caption{Plot of the function $\psi(z_i)=3\sin(2z_i)+10\xi \mathbb{I}_{(0,0.1)}(z_i) +\xi \mathbb{I}_{(0.1,\infty)}(z_i)$ for 
		three values of $\xi$.\label{sim3fig}}
\end{figure}

In this section, we consider the following left-censored PLR model with censoring levels 10\%, 20\%,
\[
Y_i = 1+3x_{1i} + \psi(z_i) + \epsilon_i, \qquad i=1,\ldots,200,
\]
where the nonparametric component $\psi(z_i)$ has the form $3\sin(2z_i)+10\xi \mathbb{I}_{(0,0.1)}(z_i) +\xi \mathbb{I}_{(0.1,\infty)}(z_i),$ in which
$\xi$ is set to vary among 0, 3 and 6. Figure \ref{sim3fig} shows the plot of $\psi(z)$ in which the jump in the function 
for different values of $\xi$ can be observed.
We assume that the realizations $(x_{1i}, z_i)$ are jointly generated 
from a bivariate normal distribution with mean zero, variance one and correlation coefficient $\rho=0.5$. The error term 
$\epsilon_i$ is also drawn from a standard normal model. In this experiment, we are interested in predicting the censored 
observations, $y_i^c$, through computing the expectation $\hat{y}_i^c=E(Y|w_i,\rho_i, \hat{\bm\Theta}) $ at the last 
iteration of the ECME algorithm. Moreover, to check the influence of noise points in model performance and imputation of 
censored observations, some noisy points simulated from $\mathcal{U}(-5,5)$, $\mathcal{U}(-3,2)$, and $\mathcal{U}(-2,8)$ 
are added, respectively to $\bm y$ and $x_1$ and $z$.

\begin{landscape}
	\begin{table}[!t]
		\begin{center}
			\caption{Performance of special cases of PLR-SMN-IC model fitted to 200 simulated datasets from either
				PLR-L-IC, PLR-SBS-IC or PLR-SGIG-IC models. \label{Tab2}}
			\hspace{-0.7cm}\scalebox{0.96}
			{\begin{tabular}{ccccccccccccccccccccc}
					\hline
					&&&&& \multicolumn{2}{c}{PLR-N-IC} && \multicolumn{2}{c}{PLR-T-IC} 
					&& \multicolumn{2}{c}{PLR-CN-IC} && \multicolumn{2}{c}{PLR-SL-IC}\\
					\cline{6-7}\cline{9-10}\cline{12-13}\cline{15-16}
					True model &Cens.   && Criterion & knots loc. & $m_1$ & $m_2$ && $m_1$ & $m_2$ && $m_1$ & $m_2$ && $m_1$ & $m_2$\\
					\hline
					&		7.5\%	&& AIC  & ES & 1609.446 & 1605.538 && 1566.020 & 1562.102 && 1573.886 & 1562.812 && 1574.804 & 1569.974\\
					&&&     & ESQ& 1609.425 & 1605.473 && 1566.202 & $\bm{1561.543}$ && 1572.966 & 1562.212 && 1574.695 & 1569.576\\
					&&& BIC & ES & 1673.310 & 1653.435 && 1637.867 & 1617.983 && 1645.733 & 1618.693 && 1642.659 & 1621.863\\
					&&&     & ESQ& 1673.288 & 1653.371 && 1638.048 & $\bm{1617.424}$ && 1644.812 & 1618.092 && 1642.550 & 1621.465\\					
					PLR-L-IC&	15\%	&& AIC  & ES & 1478.256 & 1472.341 && 1442.978 & 1438.575 && 1459.893 & 1443.115 && 1455.927 & 1449.096\\
					&&&     & ESQ& 1477.029 & 1472.298 && 1442.344 & $\bm{1438.136}$ && 1458.009 & 1442.688 && 1455.056 & 1448.503\\
					&&& BIC & ES & 1542.119 & 1520.238 && 1514.824 & 1494.455 && 1531.739 & 1498.995 && 1523.782 & 1500.985\\
					&&&     & ESQ& 1540.893 & 1520.195 && 1514.190 & $\bm{1494.017}$ && 1529.855 & 1498.569 && 1522.911 & 1500.392\\					
					&		30\%	&& AIC  & ES & 1264.806 & 1264.870 && 1233.298 & 1226.613 && 1247.416 & 1233.914 && 1243.405 & 1239.220\\
					&&&     & ESQ& 1265.242 & 1264.150 && 1231.990 & $\bm{1226.302}$ && 1247.435 & 1233.099 && 1244.752 & 1239.075\\
					&&& BIC & ES & 1328.669 & 1312.767 && 1305.144 & 1282.493 && 1319.262 & 1289.795 && 1311.260 & 1291.109\\
					&&&     & ESQ& 1329.106 & 1312.048 && 1303.836 & $\bm{1282.182}$ && 1319.282 & 1288.979 && 1312.606 & 1290.964\\ 
					\hline
					&		7.5\%	&& AIC  & ES & 1723.514 & 1719.053 && 1587.860 & 1583.466 && 1590.737 & $\bm{1583.011}$ && 1599.352 & 1595.161\\
					&&&     & ESQ& 1723.035 & 1719.418 && 1587.746 & 1583.504 && 1590.390 & 1583.220 && 1599.928 & 1594.730\\
					&&& BIC & ES & 1787.378 & 1766.951 && 1659.706 & 1639.346 && 1662.583 & $\bm{1638.892}$ && 1667.207 & 1647.050\\
					&&&     & ESQ& 1786.899 & 1767.315 && 1659.592 & 1639.384 && 1662.236 & 1639.101 && 1667.783 & 1646.619\\
					PLR-SBS-IC&	15\%	&&  AIC & ES & 1584.004 & 1580.361 && 1456.226 & $\bm{1451.544}$ && 1467.929 & 1457.499 && 1471.548 & 1465.698\\
					&&&     & ESQ& 1584.342 & 1581.378 && 1455.850 & 1451.832 && 1467.393 & 1458.350 && 1472.639 & 1467.302\\
					&&& BIC & ES & 1647.867 & 1628.258 && 1528.072 & $\bm{1507.425}$ && 1539.775 & 1513.379 && 1539.403 & 1517.587\\
					&&&     & ESQ& 1648.205 & 1629.276 && 1527.696 & 1507.713 && 1539.239 & 1514.231 && 1540.494 & 1519.191\\					
					&		30\%	&&  AIC & ES & 1399.779 & 1399.677 && 1275.221 & $\bm{1267.728}$ && 1288.915 & 1277.235 && 1291.081 & 1285.665\\
					&&&     & ESQ& 1399.622 & 1400.206 && 1273.155 & 1267.096 && 1288.500 & 1277.403 && 1291.544 & 1284.587\\
					&&& BIC & ES & 1463.643 & 1447.575 && 1347.067 & $\bm{1323.609}$ && 1360.761 & 1333.115 && 1358.936 & 1337.554\\
					&&&     & ESQ& 1463.486 & 1448.103 && 1345.001 & 1322.976 && 1360.346 & 1333.283 && 1359.399 & 1336.476\\
					\hline
					&		7.5\%	&&  AIC & ES & 1736.578 & 1734.462 && 1723.280 & $\bm{1720.268}$ && 1726.328 & 1722.382 && 1725.865 & 1722.808\\
					&&&     & ESQ& 1737.991 & 1734.704 && 1723.561 & 1720.301 && 1727.343 & 1722.575 && 1726.930 & 1722.964\\
					&&& BIC & ES & 1800.442 & 1782.359 && 1795.126 & $\bm{1776.148}$ && 1798.174 & 1778.263 && 1793.720 & 1774.697\\
					&&&     & ESQ& 1801.854 & 1782.602 && 1795.408 & 1776.182 && 1799.189 & 1778.456 && 1794.785 & 1774.854\\ 
					PLR-SGIG-IC&15\%	&&  AIC & ES & 1609.567 & 1606.316 && 1601.713 & 1597.789 && 1607.253 & 1600.947 && 1604.824 & 1600.458\\
					&&&     & ESQ& 1609.403 & 1605.598 && 1601.513 & $\bm{1597.056}$ && 1606.890 & 1600.087 && 1604.302 & 1599.547\\
					&&& BIC & ES & 1673.430 & 1654.213 && 1673.560 & 1653.670 && 1679.099 & 1656.828 && 1672.679 & 1652.347\\
					&&&     & ESQ& 1673.267 & 1653.496 && 1673.359 & $\bm{1652.937}$ && 1678.736 & 1655.968 && 1672.157 & 1651.436\\			
					&		30\%	&&  AIC & ES & 1393.320 & 1389.093 && 1388.290 & $\bm{1379.569}$ && 1395.803 & 1383.915 && 1390.980 & 1382.626\\
					&&&     & ESQ& 1394.001 & 1391.994 && 1387.456 & 1380.693 && 1395.832 & 1385.663 && 1391.137 & 1385.124\\
					&&& BIC & ES & 1457.184 & 1436.990 && 1460.136 & $\bm{1435.449}$ && 1467.676 & 1439.795 && 1458.835 & 1434.515\\
					&&&    & BIC & 1457.864 & 1439.892 && 1459.302 & 1436.574 && 1467.678 & 1441.544 && 1458.992 & 1437.013\\
					\hline
			\end{tabular}}
		\end{center}
	\end{table}
\end{landscape}

\begin{table}[!t]
	\centering
	\caption{Model performance and evaluation of the prediction accuracy for the PLR-N-IC, PLR-T-IC, PLR-CN-IC and PLR-SL-IC
		models with different censoring and noise points levels. \label{Tab3}}
	\hspace{-0.3cm}\scalebox{0.9}
	{\begin{tabular}{lcccccccccccccccccccc}
			\hline
			&&& \multicolumn{2}{c}{PLR-N-IC} && \multicolumn{2}{c}{PLR-T-IC} 
			&& \multicolumn{2}{c}{PLR-CN-IC} && \multicolumn{2}{c}{PLR-SL-IC}\\
			\cline{4-5}\cline{7-8}\cline{10-11}\cline{13-14}
			Cens.   &Noise& $\xi$ & BIC & MAE && BIC & MAE && BIC & MAE && BIC & MAE\\
			\hline
			&0      & 0  & 556.012 & 0.669 && 566.464 & 0.674 && 566.410 & 0.671 && 561.050 & 0.671\\
			&& 3 & 555.444 & 0.669 && 565.887 & 0.672 && 565.849 & 0.669 && 560.449 & 0.669\\
			&& 6 & 555.125 & 0.668 && 565.605 & 0.671 && 565.593 & 0.668 && 560.209 & 0.668\\
			\\
			10\% &10       & 0 & 732.298 & 0.671 && 669.805 & 0.654 && 666.365 & 0.654 && 666.271 & 0.658\\
			&& 3 & 762.192 & 0.687 && 679.711 & 0.666 && 673.627 & 0.666 && 673.007 & 0.665\\
			&& 6 & 789.640 & 0.707 && 689.772 & 0.671 && 683.301 & 0.672 && 682.814 & 0.671\\
			\\
			&20        & 0 & 860.391 & 0.730 && 763.462 & 0.679 && 757.945 & 0.678 && 759.861 & 0.685\\
			&& 3 & 899.970 & 0.753 && 777.642 & 0.699 && 772.075 & 0.696 && 771.062 & 0.709\\
			&& 6 & 920.889 & 0.786 && 787.971 & 0.712 && 782.039 & 0.704 && 781.512 & 0.728\\	
			\hline
			& 0      & 0 & 503.553 & 0.706 && 514.279 & 0.719 && 514.320 & 0.708 && 508.880 & 0.711\\
			&	& 3 & 504.768 & 0.709 && 515.308 & 0.714 && 515.196 & 0.710 && 509.656 & 0.709\\
			&& 6 & 504.169 & 0.704 && 514.603 & 0.707 && 514.538 & 0.705 && 509.011 & 0.704\\
			\\
			20\%  &10     & 0  & 682.773 & 0.744 && 617.020 & 0.707 && 615.887 & 0.695 && 614.687 & 0.704\\
			&& 3 & 711.959 & 0.687 && 628.295 & 0.666 && 625.234 & 0.667 && 621.523 & 0.665\\
			&& 6 & 738.295 & 0.767 && 637.143 & 0.709 && 632.903 & 0.702 && 630.251 & 0.704\\
			\\
			&20        & 0  & 787.876 & 0.801 && 700.524 & 0.716 && 695.183 & 0.706 && 697.132 & 0.717\\
			&& 3 & 825.702 & 0.810 && 714.499 & 0.732 && 710.151 & 0.724 && 707.972 & 0.733\\
			&& 6 & 853.141 & 0.853 && 726.726 & 0.745 && 721.391 & 0.731 && 720.359 & 0.749\\			
			\hline
	\end{tabular}}
\end{table}
For each of the 200 generated samples of size 200, we fit the PLR-N-IC, PLR-T-IC, PLR-CN-IC and PLR-SL-IC models to the
data by considering $m_2$ and ESQ approaches for number of knots and their positions, as well as their values of BIC are recorded.
Following \cite{Matos2016}, to investigate the performance of the prediction of censored observations, we compute the 
mean absolute error (MAE) defined as
\[
\text{MAE}=\frac{1}{n_c200}\sum_{j=1}^{200}\sum_{i=1}^{n_{c}}|\hat{y}_{ij}-{y}_{ij}|,
\]
where ${y}_{ij}$ and $\hat{y}_{ij}$ are the actual and predicted values of the $i$th realization in the $j$th trail
and $n_{c}$ denotes the number of censored observations. Table \ref{Tab3} shows the average values of BIC and MAE over
200 replications for different levels of censoring, values of $\xi$, and number of noise points. The results depicted in 
Table \ref{Tab3} indicate that the value of MAE increases as the number of censored observations is increased.
As can be expected, the PLR-N-IC model preforms well when the number of noise realizations is zero. However, adding 
the noise points reduces its flexibility in both fitting data and predicting censored observations. 
It can be observed that only the MAE of the PLR-N-IC model increases as the noise points are added to the datasets, 
showing its lack of robustness in dealing with the contaminated data with the noise observations.
One can see from the values of BIC, the PLR-SL-IC model is the best model for all scenarios with noise observations.

\subsection{Robustness of the EM estimates}
In this Monte-Carlo experiment, we are interested in comparing the performance of the parameter estimates in the presence of 
outliers on the response variable. We simulate 500 Monte Carlo samples from the interval-censored PLR model \eqref{PLR} 
of size $n = 300$, while the errors are randomly generated from the normal distribution with scale parameter $\sigma^2=2$.
The imposed censoring levels are 10\%, 20\% and 30\%. We set $\bm\beta^\top=(1,4,2)$ and generate the covariates 
$\bm x_i^\top=(1,x_{1i},x_{2i})$ independently from $x_{1i}\sim\mathcal{U}(2,20)$ and $x_{2i}\sim\mathcal{B}(1,0.6)$.
The nonparametric part of the model is also generated by the function $\psi(z_i)=\cos(4\pi z_i)\exp(-z_i^2/2),$ where 
$z_{i}\sim\mathcal{U}(0, 3)$. Furthermore, we add perturbations into the largest uncensored observation, namely $y_{\max}^* 
= y_{\max}-\delta $ with $\delta$ varying from 1 to 10. The parameter estimates are computed under the PLR-N-IC, PLR-T-IC,
PLR-SL-IC and PLR-CN-IC models in each replication with and without contaminations, denoted by $\hat{\theta}_{(\delta)}$ 
and $\hat{\theta}$, respectively. We use $m_2$ and ESQ approaches for number of knots and their positions.

\begin{figure}[!t]
	\centerline{\includegraphics[height=6cm, width=5cm]{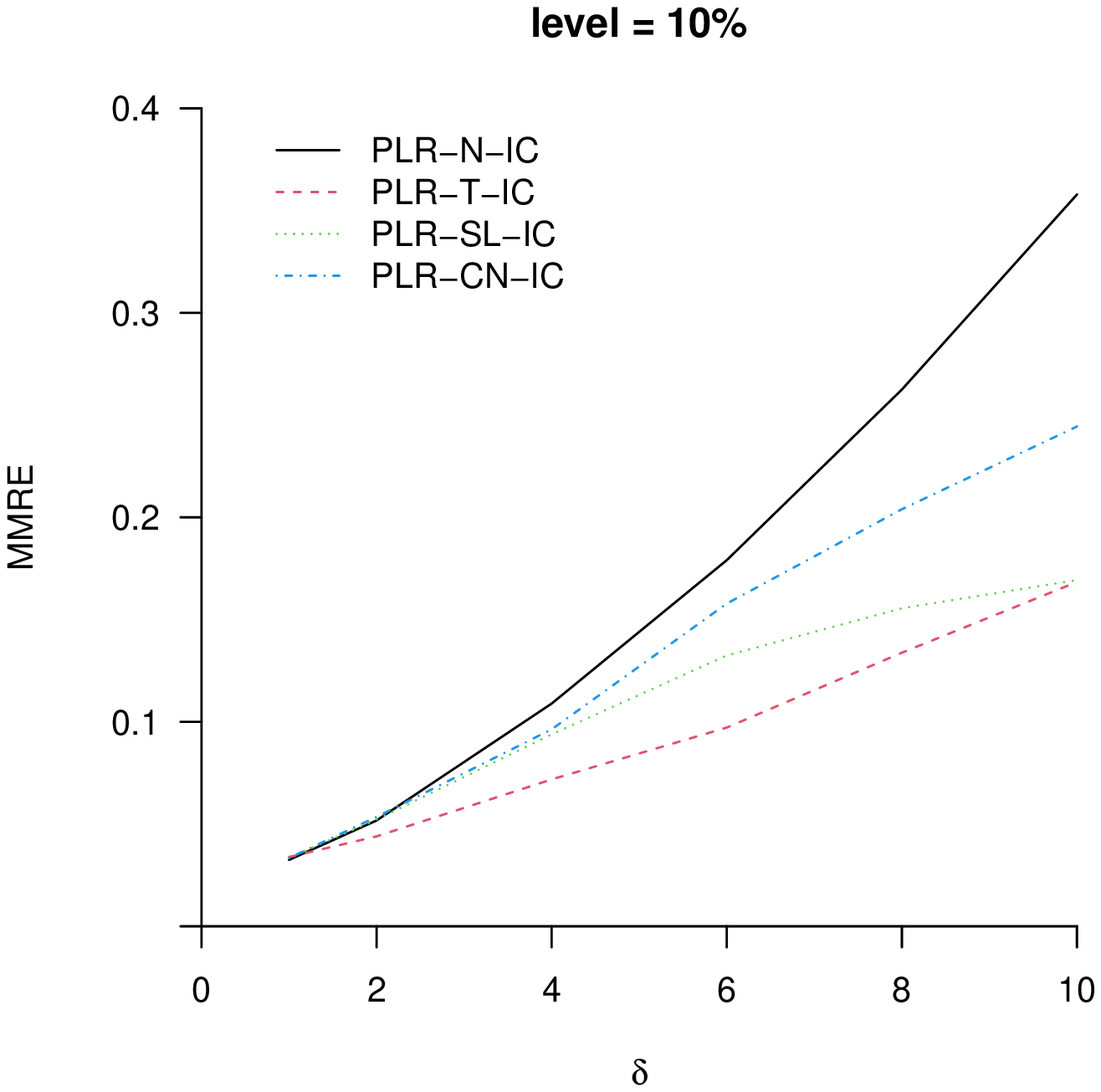}
		\includegraphics[height=6cm, width=5cm]{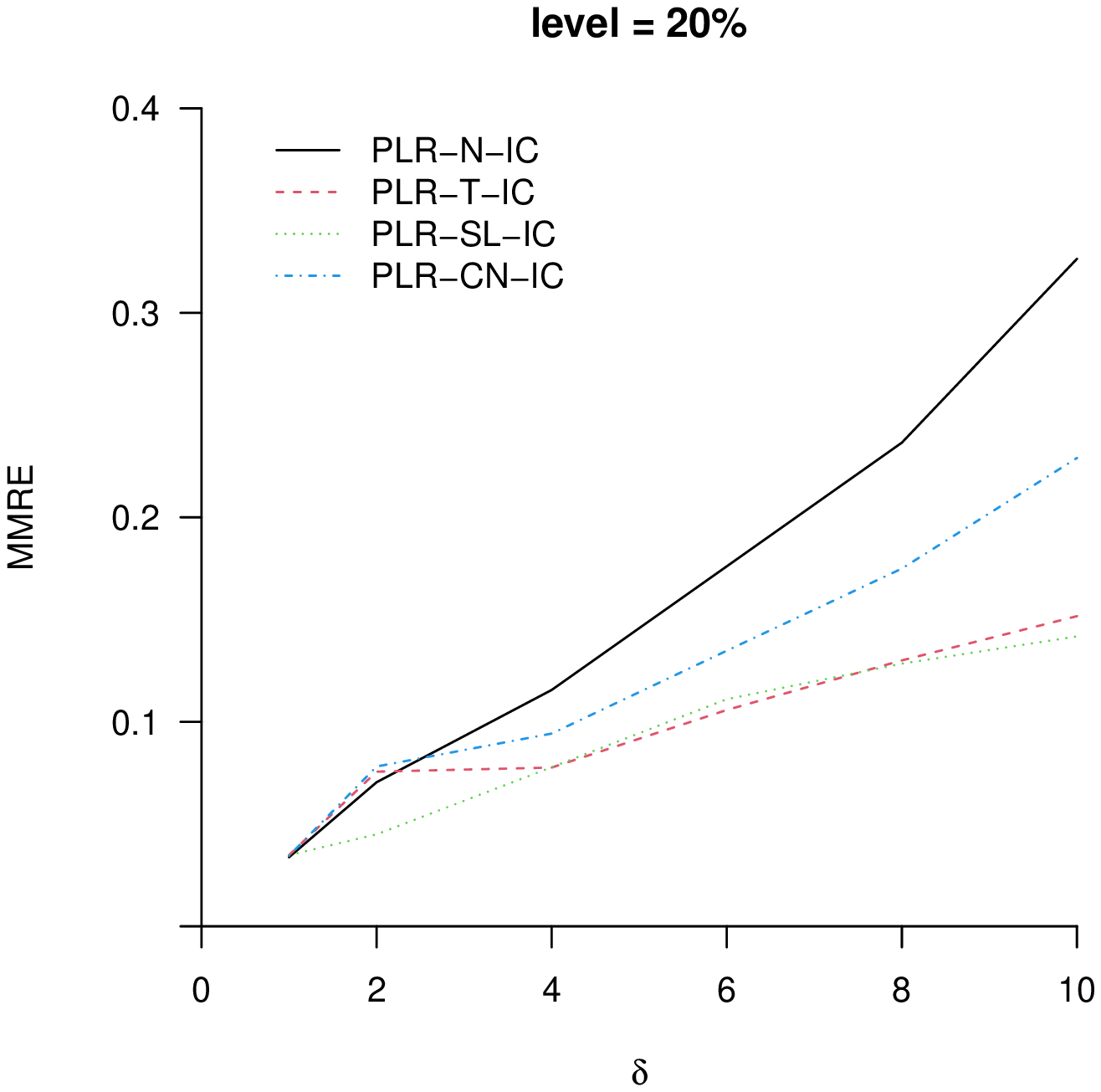}
		\includegraphics[height=6cm, width=5cm]{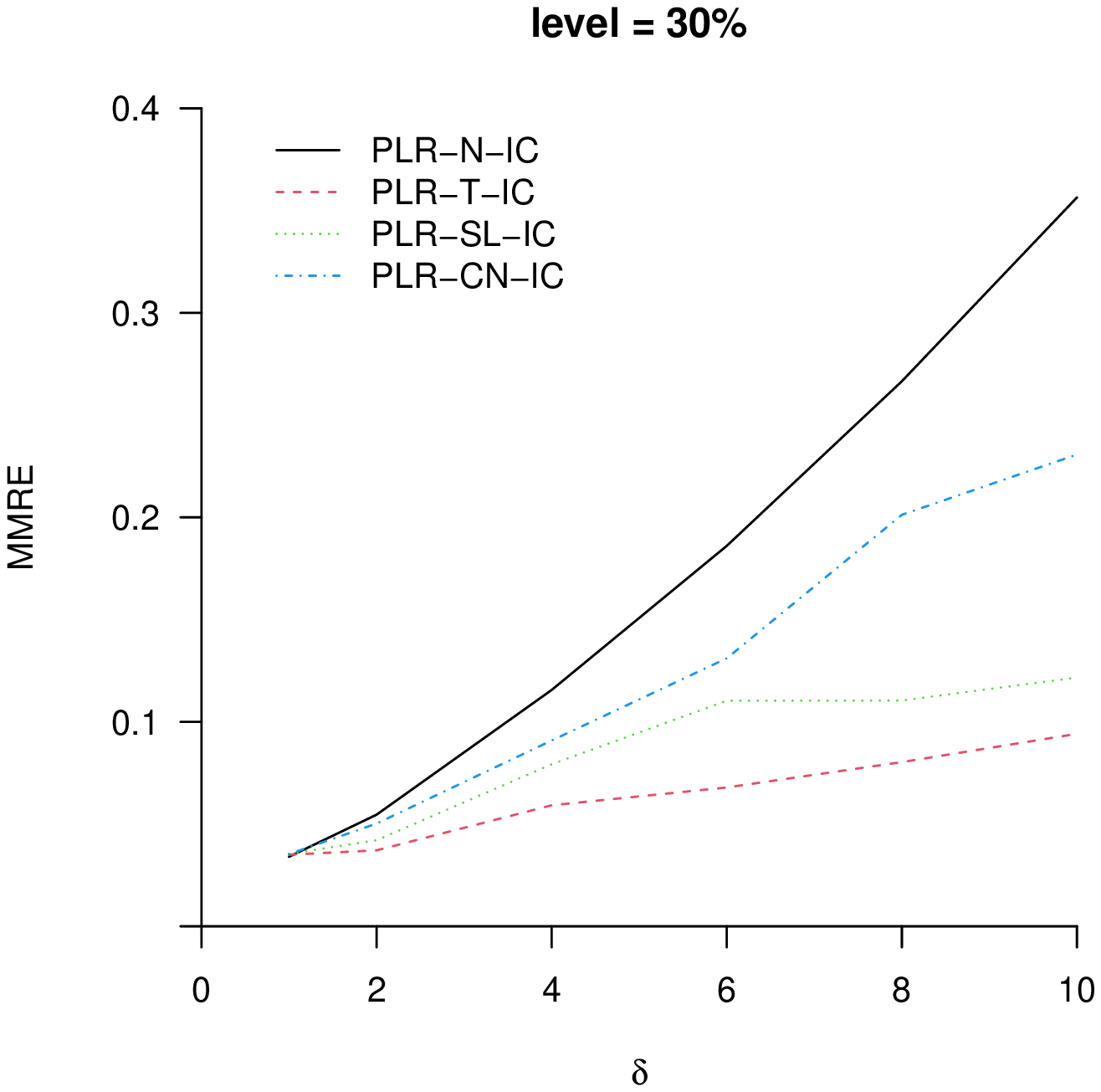}}
	\caption{Average MMRE on estimates for different contaminations $\delta$ and three censoring levels: 10\%, 20\% and 30\%.\label{sim4fig}}
\end{figure}

To assess relative changes on the parameter estimates by the presence of outliers, the mean magnitude of relative error (MMRE;
\cite{fagundes2013}), defined is calculated as follows:
\[
\text{MMRE} =\frac{1}{2000}\sum_{l=1}^{500}\left\lbrace\sum_{j=1}^3\left( \frac{\hat{\beta}_{jl(\delta)}-
	\hat{\beta}_{jl}}{\hat{\beta}_{jl}}\right) +\frac{\hat{\sigma}_{(\delta)}^2- \hat{\sigma}^2}{\hat{\sigma}^2}\right\rbrace.  
\]
Curves of the average MMRE as a function of contamination level $\delta$ are displayed in Figure \ref{sim4fig}. 
It could be seen that the influence of the outliers in parameter estimation increases as $\delta$ increases for all models. 
As one would expect, the heavy-tailed models, such as PLR-T-IC and PLR-SL-IC, are less adversely affected, showing
their robustness against the presence of outliers. On the other hand, an extreme observation
seems to be much more effective on the PLR-N-IC, reflecting a lack of ability to reduce the influence of outliers.

\section{Real data analyses}\label{sec5}
In the following, we present two applications of the PLR-SMN-IC model to real data for illustrative purposes. 
The first real dataset corresponds to the young married women's labor force participation using the data extracted from the 
Canadian Survey of Labour and Income Dynamics (SLID). The complete data, reported in \cite{fox2015}, consists of 6900 
respondents of the married women aging from 18 and 65 years. However, the samples with missing data (near to 7\%) are removed 
from the study. For each 6340 women in the remaining sample, four measures, namely working hours, family income, age and education, are recorded.
By way of illustration, we consider the PLR model as
\[
y = \beta_0+\beta_1x_1+\beta_2x_2+\psi(z) + \epsilon,
\]
where $y$ is the working hours scaled by 1000, $x_1$ age, $x_2$ family income scaled by 1000 and $z$ as education. 
We note that among 6340 women in the sample, only 4398 represent propensity to work outside the home, i.e. if that propensity is 
above the threshold 0, we observe positive hours worked. Therefore, the response variable is left-censored at the value 0.

By way of the second illustration, we consider extramarital Affairs data that is available on the ``$\textbf{AER}$" packages of $\texttt{R}$. 
The Affairs dataset, originally reported in \cite{fair1978}, is recently re-analyzed by \cite{greene2003} in the censored regression framework.
The variables involved in the study of the Affairs dataset were: the number of observations engaged in extramarital sexual intercourse during the past year (y),
number of years married ($x_1$), occupation according to Hollingshead classification ($x_2$), self rating of marriage ($x_3$) and
the age as the nonparametric component $z$. Recommended by \cite{greene2003}, the response variable is highly left-censored at zero (451 out of 601, or 75\%)
which may have significant effects on the coefficients.

\begin{figure}[!t]
	\centerline{\includegraphics[height=6cm, width=7cm]{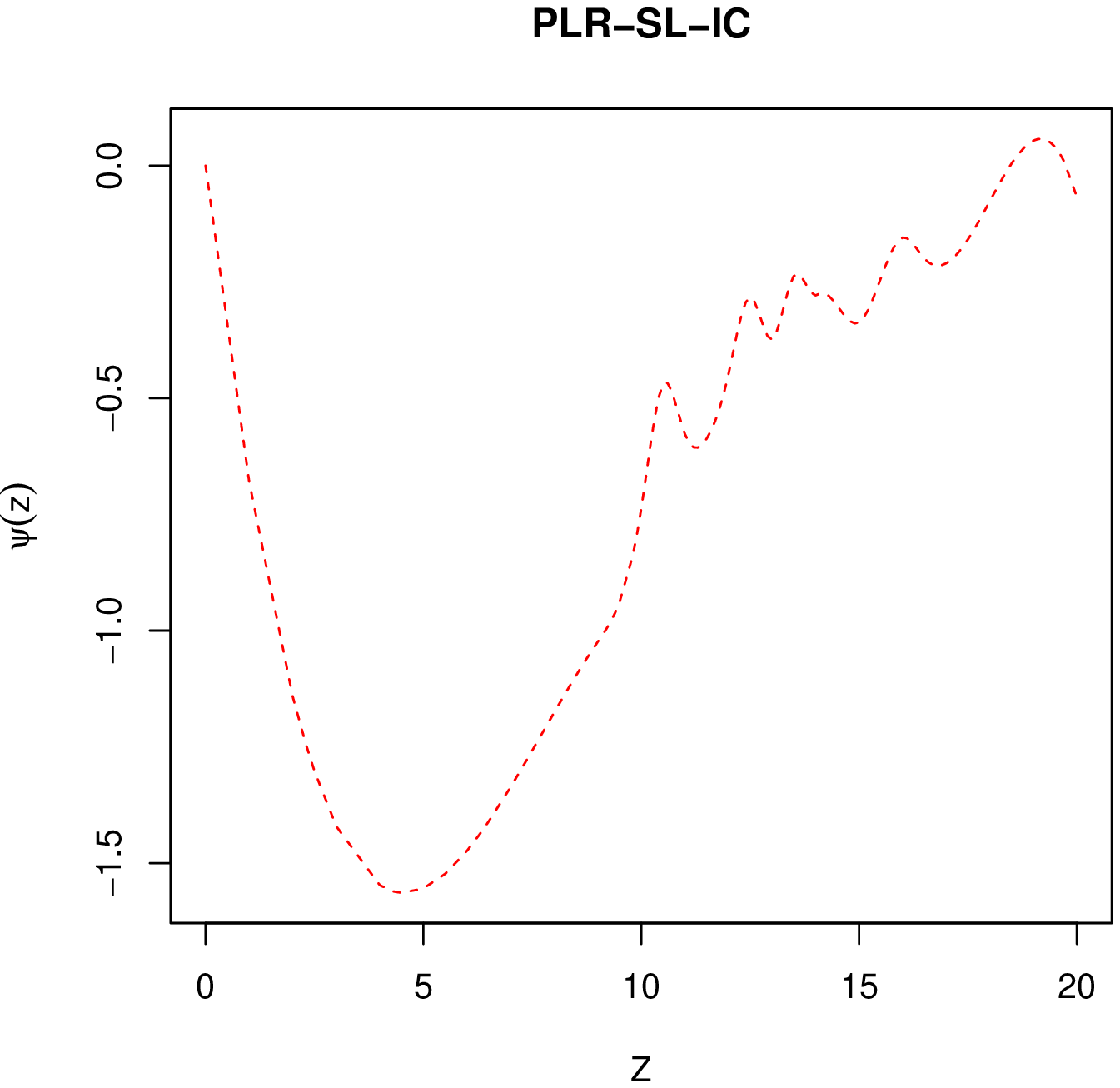}
		\includegraphics[height=6cm, width=7cm]{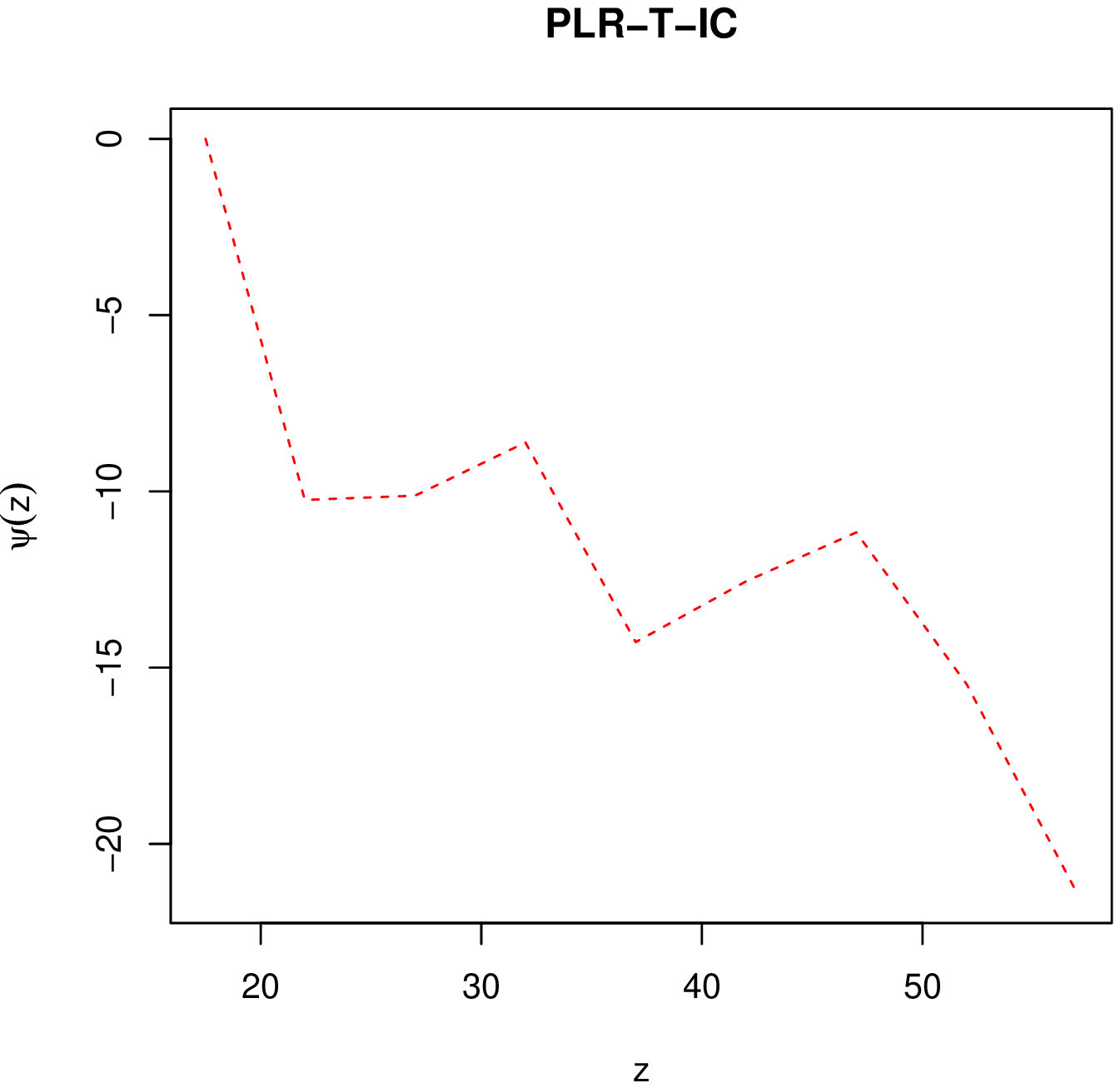}}
	\caption{The curves of the estimated unknown $\psi(z)$ for the best fitted model to the SLID (left panel) and Affairs (right panel)
		datasets.\label{Real3}}
\end{figure}

\begin{table}[!t]
	\centering
	\caption{Model performance and evaluation of the prediction accuracy for the PLR-N-IC, PLR-T-IC, PLR-CN-IC and PLR-SL-IC
		models. \label{Real1}}
	\scalebox{0.85}
	{\begin{tabular}{lcccccccccccccccccccc}
			\hline
			&& \multicolumn{2}{c}{PLR-N-IC} && \multicolumn{2}{c}{PLR-T-IC} 
			&& \multicolumn{2}{c}{PLR-CN-IC} && \multicolumn{2}{c}{PLR-SL-IC}\\
			\cline{3-4}\cline{6-7}\cline{9-10}\cline{12-13}
			Data & Criterion & $m_1$ & $m_2$ && $m_1$ & $m_2$ && $m_1$ & $m_2$ && $m_1$ & $m_2$\\
			\hline
			SLID & $\ell_{\max}$ & -7990.325 & -8082.835 && -7888.270 & -8030.143 && -7941.59 & -8043.783 && $\bm{-7879.242}$ & -8015.184 \\ 
			& AIC & 16022.650 & 16191.670 && 15820.540 & 16088.290 && 15929.180 & 16117.570 && $\bm{15802.480}$ & 16058.370 \\ 
			& BIC & 16164.500 & 16279.480 && 15969.140 & 16182.850 && 16084.540 & 16218.890 && $\bm{15951.090}$ & 16152.930 \\ 
			Affairs & $\ell_{\max}$ & -689.7863 & -693.7968 && -678.1139 & $\bm{-671.7741}$ && -680.557 & -686.215 && -675.985 & -679.5878 \\ 
			& AIC & 1407.573 & 1411.594 && 1386.228 & $\bm{1369.548}$ && 1393.114 & 1400.430 && 1381.970 & 1385.176 \\ 
			& BIC & 1469.153 & 1464.377 && 1452.207 & $\bm{1426.730}$ && 1463.492 & 1462.011 && 1447.949 & 1442.357\\ 		
			\hline
	\end{tabular}}
\end{table}
We fit the PLR-N-IC, PLR-T-IC, PLR-CN-IC and PLR-SL-IC models to each of the datasets by implementing the proposed ECME algorithm in Section
\ref{sec4}. We use both $m_1$ and $m_2$ approaches for choosing the number of interior knots. Exploiting the ES and ESQ methods for 
the location of the knots suggest that the ESQ has better performance for analyzing these datasets. 
Table \ref{Real1} presents the maximized log-likelihood function, and the values of AIC and BIC for all fitted
models with both $m_1$ and $m_2$ number of interior knots. It can be observed that the PLR-SMN-IC models with heavy tails
have better fit than the PLR-N-IC model. Results based on AIC and BIC indicate that the PLR-SL-IC and PLR-T-IC models
provide a highly improved fit of the data over other models, for the SLID and Affairs datasets respectively.
Table \ref{Real2} shows the parameter estimates of the best fitted PLR-SMN-IC sub-models with respect to the number of knots,
for the SLID and Affairs datasets. It can be seen that all models offer similar estimates for the slope parameters ($\beta_1,\beta_2$ and $\beta_3$).
Moreover, the parameter estimate $\bm\nu$ is significantly different from zero and has large values for the PLR-T-IC, PLR-CN-IC, 
and PLR-SL-IC models, indicating the departure of the data from the normality assumption.
Finally, we display in Figure \ref{Real3} the estimated curve of the nonparametric component $\psi(z)$ for the fitted PLR-SL-IC and 
PLR-T-IC models to the SLID and Affairs datasets, respectively.

\begin{table}[!t]
	\centering
	\caption{ML parameter estimates of the PLR-SMN-IC models for two considered datasets.\label{Real2}}
	\scalebox{0.85}
	{\begin{tabular}{cccccccccccccccccccc}
			\hline
			&& \multicolumn{2}{c}{PLR-N-IC} && \multicolumn{2}{c}{PLR-T-IC} 
			&& \multicolumn{2}{c}{PLR-CN-IC} && \multicolumn{2}{c}{PLR-SL-IC}\\
			\cline{3-4}\cline{6-7}\cline{9-10}\cline{12-13}
			parameter && SLID & Affairs && SLID & Affairs && SLID & Affairs && SLID & Affairs\\
			\hline
			$\beta_0$ && 2.1600 & 11.0125&& 2.2760 & 13.3676&& 2.1492 & 7.7550 && 2.2361 & 13.3150 \\
			$\beta_1$ && -0.0180& 0.3021 && -0.0179& 0.3453 && -0.0173& 0.3030 && -0.0177& 0.3606\\
			$\beta_2$ && -0.0031& 0.3432 && -0.0028& 0.2699 && -0.0030& 0.2545 && -0.0027& 0.3085\\
			$\beta_3$ && -- & -2.7982 && -- & -3.0777 && -- & -1.9817 && -- & -3.1318\\
			$\sigma$  && 1.1719 & 8.9666 && 0.9816 & 6.7422 && 0.8340 & 4.2330 && 0.8155 & 6.0719\\
			$\nu$     && -- & -- && 6.8342 & 8.0434 && 0.9661 & 0.4432 && 2.8531 & 2.1765\\
			$\gamma$  && -- & -- && -- & -- && 0.6380 & 0.2355 && -- & -- \\		
			\hline
	\end{tabular}}
\end{table}

\section{Conclusion}\label{sec6}
This paper proposed a semiparametric inference for partially linear regression model with interval-censored responses where the SMN family of distributions is assumed for the error term. The new model provided an
alternative benchmark for the conventional choice of the normal distribution. Our proposed model extended the recent works by \cite{Garay2015,garay2016} to the 
partially linear regression framework that, to the best of our knowledge, there are no previous studies of a likelihood-based perspective
related to this topic. 
Using the basis spline function, B-spline, we defined the pseudo covariates and pseudo regression parameter. We then exploited
the hierarchical representation of the SMN class of distributions for developing a feasible ECME algorithm to obtain the ML parameter estimates.

We conducted four simulation studies to examine the performance of the proposed model and its parameter estimation. Specifically, simulation studies
aimed at checking the ability of the new methodology in parameter recovery, model comparisons and sensitivity analysis in presence of noise points and
a single outlier. Finally, two real-world datasets, SLID and Affairs, were analyzed for illustrative purposes.  
As was reported, the heavy-tailed PLR-SMN-IC models, such as PLR-T-IC, PLR-CN-IC and PLR-SL-IC, presented better results than the PLR-N-IC model to these 
real data examples. All computations were carried out by $\texttt{R}$ language and the computer program is available from the 
first author upon request. 

The methodology presented in this paper can be extended through the following open issues:
\begin{itemize}
	\item[$\bullet$] Motivated from the SLID dataset, it is interesting to formulate a PLR model to simultaneously handle missing and censored 
	observations \cite{Lin2018, Lin2019}.
	
	\item[$\bullet$] Motivated from the Affairs dataset and recommended by \cite{greene2003}, one can construct a PLR-SMN model for analyzing 
	doubly-censored data.
	
	\item[$\bullet$] Two important questions that might be raised are: how can we handle multivariate covariates in the nonparametric part? and 
	how can we select the vector of explanatory covariates for both linear and nonparametric parts?. The former issue can be addressed through either
	the single-index model, generalized additive model or multivariate B-spline function, whereas the latter can be answered by variable selection 
	studies.  
	
	\item[$\bullet$] As future research, we are exploring in building a finite mixture of semiparametric partially linear regression models as 
	an extension of \cite{Zeller2018, mirfarah2020} in both likelihood and Bayesian context \cite{Lin2004, Lin2011}.
\end{itemize}

\section*{Acknowledgments}
	This work is based upon research supported by the South Africa National Research Foundation and South Africa
	Medical Research Council (South Africa DST-NRF-SAMRC SARChI Research Chair in Biostatistics, Grant number
	114613), as well as by the National Research Foundation of South Africa (Grant Numbers 127727). Opinions
	expressed and conclusions arrived at are those of the author and are not necessarily to be attributed to the NRF.

\section*{Conflict of interest}
The authors declare that they have no conflict of interest.
\appendix
\section*{Appendix A: Conditional expectations of the special cases of the SMN distributions}\label{appa}
\def\theequation{A.\arabic{equation}}
\setcounter{equation}{0}
{\bf Uncensored observations:} 
For the uncensored data $y$, we have $\rho =0$. Therefore, the only necessary conditional expectation
$\hat{u}=E(U|Y=y,\hat{\bm\Theta})$ for the considered models can be computed as follows. 

\begin{itemize}
	\item[$\bullet$] If $Y\sim \mathcal{N}\big(\mu,\sigma^2\big)$, in this case, $U=1$ with probability one, 
	and so $\hat{u}=1$.
	
	\item[$\bullet$] If $Y\sim \mathcal{T}\big(\mu,\sigma^2,{\nu}\big)$, we have
	\begin{equation*}
	\hat{u}= \frac{\hat{\nu} + 1}{\hat{\nu} +\delta\Big(y, \hat{\mu},\hat{\sigma}\Big) },
	\end{equation*}
	where $\delta(y,\mu,\sigma) =\Big((y-\mu)/\sigma \Big)^2$.
	
	\item[$\bullet$] If $Y\sim \mathcal{SL}\big(\mu,\sigma^2,{\nu}\big)$, then
	\begin{equation*}
	\hat{u}= 2 \left(\delta\Big(y, \hat{\mu},\hat{\sigma}\Big)\right)^{-1} 
	\frac{\Gamma\left(\hat{\nu} + 1.5,\; 0.5 
		\delta\Big(y, \hat{\mu},\hat{\sigma}\Big)\right) }{
		\Gamma\left(\hat{\nu}+ 0.5,\; 0.5  
		\delta\Big(y, \hat{\mu},\hat{\sigma}\Big)\right)}.
	\end{equation*}
	
	\item[$\bullet$] If $Y\sim \mathcal{CN}\big({\mu},{\sigma}^2,{\nu},{\gamma}\big)$, we have
	\begin{equation*}
	\hat{u}= \frac{1-\hat{\nu} + \hat{\nu}\hat{\gamma}^{1.5} \exp\Big\{0.5(1-\hat{\gamma}) 
		\delta\Big(y, \hat{\mu},\hat{\sigma}\Big)\Big\}}{
		1-\hat{\nu} + \hat{\nu}\hat{\gamma}^{0.5} 
		\exp\Big\{0.5(1-\hat{\gamma}) \delta\Big(y, \hat{\mu},\hat{\sigma}\Big)\Big\}}.
	\end{equation*}
	
\end{itemize}

{\bf Censored cases:} 
In the censored cases, we have $\rho=1$. For the sake of notation, 
let 
$$T = \dfrac{Y- \hat{\mu}}{\hat{\sigma}}\sim\text{SMN}(0,1,\hat{\bm\nu}), \qquad 
\hat{t}_1=\frac{c_{1}- \hat{\mu}}{\hat{\sigma}}, \qquad
\hat{t}_2=\frac{c_{2}- \hat{\mu}}{\hat{\sigma}}.  $$
Therefore, the necessary conditional expectations 
$\hat{u}=E(U|c_{1} \leq Y \leq c_{2},\hat{\bm\Theta})$,
$\widehat{uy}= E( UY | c_{1} \leq Y \leq c_{2}, \hat{\bm\Theta})$, and
$\widehat{uy^2} = E( UY^2 | c_{1} \leq Y \leq c_{2}, \hat{\bm\Theta})$
for the considered special models can be computed as follows.
\begin{align*}
\hat{u} &= E\left(U|\hat{t}_{1}\leq T \leq \hat{t}_{2}, \hat{\bm\Theta}\right) 
= \frac{E_\Phi\left(1,\hat{t}_{2}\right) - E_\Phi\left(1,\hat{t}_{1}\right)}{F_{SMN}\left( 
	\hat{t}_{2};\hat{\bm\nu}\right)-F_{SMN}\left( \hat{t}_{1};\hat{\bm\nu}\right) },\\
\widehat{uy} &= \hat{\mu} \hat{u} + 
\hat{\sigma}_j^{(k)}  E\left(U T\Big| \hat{t}_{1} \leq T\leq \hat{t}_{2}, \hat{\bm\Theta}\right)
= \hat{\mu} \hat{u}+ \hat{\sigma}_j^{(k)} \left\lbrace  \frac{E_\phi\left(0.5,\hat{t}_{1}\right) - E_\phi\left(0.5,\hat{t}_{2}\right)}{
	F_{SMN}\left(\hat{t}_{2};\hat{\bm\nu}\right)-F_{SMN}\left( \hat{t}_{1};\hat{\bm\nu}\right) } \right\rbrace ,\\
\widehat{uy^2}
&= \hat{\mu}^2 \hat{u}
+ 2\hat{\mu}\hat{\sigma} \widehat{uy}
+\hat{\sigma}^2 E\left(U T^2\Big| \hat{t}_{1} \leq T\leq \hat{t}_{2}, \hat{\bm\Theta}\right), \nonumber \\
&= \hat{\mu}^2 \hat{u} + 2\hat{\mu}\hat{\sigma} \widehat{uy} 
+ \frac{\hat{\sigma}^{2}\left( E_\Phi\left(1,\hat{t}_{2}\right) - E_\Phi\left(1,\hat{t}_{1}\right) + 
	\hat{t}_{1} E_\phi\left(0.5,\hat{t}_{1}\right)-\hat{t}_{2} E_\phi\left(0.5,\hat{t}_{2}\right)  \right)}{F_{SMN}\left(\hat{t}_{2};\hat{\bm\nu}\right)-F_{SMN}\left( \hat{t}_{1};\hat{\bm\nu}\right)}, 
\end{align*}

where
\[
E_\phi(r,h) = E\left(U^r \phi\big(h\sqrt{U}\big) \right)\quad \text{and} \quad 
E_\Phi(r,h) = E\left(U^r \Phi\big(h\sqrt{U}\big) \right).
\]
In the following, the closed forms of $E_\phi(r,h)$ and $E_\Phi(r,h)$ for the special cases of SMN class of distributions are presented.
\begin{itemize}
	\item[$\bullet$] For the normal distribution, we have
	\begin{align*}
	E_\phi(r,h) = \phi(h) \quad \text{and} \quad E_\Phi(r,h)=\Phi(h).
	\end{align*}
	
	\item[$\bullet$] In the case of Student-$t$ distribution, we have	
	\begin{align*}
	E_\phi(r,h) &= \frac{\Gamma\left(\dfrac{\hat{\nu}+2r}{2}\right) }{\sqrt{2\pi}\Gamma(\hat{\nu}/2)} 
	\left(\frac{\hat{\nu}}{2} \right)^{\dfrac{\hat{\nu}}{2}} \left(\frac{2}{h^2+\hat{\nu}} \right)^{\dfrac{\hat{\nu}+2r}{2}},\\
	E_\Phi(r,h) &= \Gamma\left( \frac{\hat{\nu}+2r}{2}\right) \left(\frac{2}{\hat{\nu}} \right)^r F_{PVII}\left(h;\hat{\nu}+2r, \hat{\nu}\right) \Big/ \Gamma(\frac{\hat{\nu}}{2}).
	\end{align*}
	where $F_{PVII}(\cdot;\nu,\delta)$ denotes the cdf of Pearson type $VII$ distribution.
	\item[$\bullet$] For the slash model, we have
	\begin{align*}
	E_\phi(r,h)= \frac{\hat{\nu}}{\sqrt{2\pi}} \left( \frac{2}{h^2} \right)^{\hat{\nu}+r}\Gamma(\hat{\nu}+r, \frac{h^2}{2} )
	\quad \text{and} \quad 
	E_\Phi(r,h) = \frac{\hat{\nu}}{\hat{\nu}+r} F_{SL}\left(h;\hat{\nu}+r \right).
	\end{align*}
	
	\item[$\bullet$] For the contaminated-normal distribution, we have
	\begin{align*}
	E_\phi(r,h) &=\left(\hat{\gamma}\right)^r \hat{\nu}\phi\left(h\sqrt{\hat{\gamma}}\right) 
	+\left(1-\hat{\nu}\right) \phi(h),\\
	E_\Phi(r,h) &=\left(\hat{\gamma}\right)^r F_{CN}\left(h;\hat{\nu},\hat{\gamma}\right)+ 
	\left(1-\left(\hat{\gamma}\right)^r\right) \left(1-\hat{\nu}\right) \Phi(h).
	\end{align*}	
\end{itemize}



\end{document}